\documentclass[prb, twocolumn, showpacs, amsmath, amssymb,superscriptaddress]{revtex4-1}

\usepackage{amssymb,amsfonts,amsmath} 
\usepackage{graphicx}
\usepackage{color}
\usepackage{hyperref}
\usepackage{longtable}
\usepackage{bbm}
\usepackage{ulem}
\begin{document} 

\title{The Adiabatically Deformed Ensemble: Engineering Non-Thermal States of Matter}

\author{D.\ M.\ Kennes}
\affiliation{Department of Physics, Columbia University, New York, NY 10027, USA}

\begin{abstract} 
We propose a route towards engineering non-thermal states of matter, which show largely unexplored physics.  The main idea relies on the adiabatic passage of a thermal ensemble under slow variations of the system Hamiltonian. If the temperature of the initial thermal ensemble is either zero or infinite the ensemble after the passage is a simple thermal one with the same vanishing or infinite temperature. However, for any finite non-zero temperature intriguing non-thermal ensembles can be achieved. We exemplify this in: (a) a single oscillator (b) a dimerized interacting one dimensional chain  of spinless fermions, (c) a BCS-type superconductor and (d) the topological Kitaev chain. We solve these models with a combination of methods; either exactly, numerically using the density matrix renormalization group (DMRG) or within an approximate functional renormalization group (FRG) scheme. The designed states show strongly non-thermal behavior in each of the considered models. For example, for the chain of spinless fermions we exemplify how long ranged non-thermal power-law correlations can be stabilized and for the Kitaev chain we elucidate how the non-thermal ensemble can largely alter the transition temperature separating topological and trivial phases. 
\end{abstract}

\pacs{} 
\date{\today} 
\maketitle

Non-equilibrium states of matter have attracted a great deal of interest lately. To achieve a steady non-equilibrium state one can, e.g., contact a system by leads to drive currents through it. Another route to non-equilibrium is to isolate a system (sufficiently well) from its environment and subsequently change its Hamiltonian. The second class includes (e.g.) pump-probe experiments \cite{Yu91,Miyano97,Wall11,Blumenstein11} or quenching  ultra cold gases \cite{Bloch08,Polkovnikov11} by abruptly tuning through a Feshbach resonance. Quantum quenches have been studied in great detail.\cite{Polkovnikov11,Yukalov11,Gogolin16,Essler16,Vidmar16}


Here, we study the adiabatic deformation of a given system. In this context, the Kibble-Zurek  mechanism \cite{KZ} has gained a lot of attention.   At the critical point of a second order phase transition the critical slowing down implies that it is impossible to drive a system adiabatically through its transition.
The finite speed of the passage disturbs the ordering of the state after the transition, leaving behind ordered domains (with size depending on the rate of change and the critical exponents).\cite{KZ} The Kibble-Zurek  mechanism has been tested extensively numerically\cite{KZnum} as well as experimentally.\cite{KZexp} 
We focus on adiabatic deformation of a given initial ensemble staying away from crossing a second order critical point. Studies of such adiabatic deformations  or in general finite time quenches (without crossing a phase boundary) mainly focus on the initial state being the ground state.\cite{Quenchfinite,Dora11} These works either cover the dynamics in the most general case in between sudden quenches or adiabatic deformations or address the important questions whether adiabatic evolution is possible at all and if so identify the leading corrections in the rate of change. This is a pressing matter, because in Ref.~\onlinecite{noadia}, it was shown that not all systems exhibit adiabatic behavior. In this work the authors separate systems into three generic classes with respect to the behavior of the excess energy under slow variations of the Hamiltonian: (a) analytic, where the corrections to the adiabatic behavior vanish as the square of the rate of change, (b) non-analytic, where the corrections vanish following a non-quadratic behavior and (c) non-adiabatic, where the corrections depend on a power-law in the system size and thus as the system size approaches infinity the adiabatic limit ceases to exist. We note that we use the term 'adiabatic' in contrast to this classification in a different convention, which is also common in the literature: adiabatic variation here denotes sufficiently slow variations of a system parameter, such that further decreasing the speed of the variation yields no changes in physical observables (supported either on all energy/inverse length scales or a subset thereof).  

A marked exception to the study of ground state properties under adiabatic evolution is listed in Ref. \onlinecite{FiniteT_quench}, where it was shown  that adiabaticity in a Luttinger liquid (at least for sufficiently smooth variations, such that the Luttinger liquid picture remains valid) falls into the analytic class of Ref~\onlinecite{noadia} when deforming the ground state by changing the two-particle interaction. For finite $T$ initial states this changes to a linear dependency of the excess energy on the rate of change. However, within this study the adiabatic deformation, even at finite temperature, amounts simply to a change in temperature of the ensemble after the deformation. This is a direct consequence of the linear dispersion of the assumed model and the studied type of deformation, i.e. slowly changing the interaction (see below).

We find that in general and  for a broad variety of quantum systems by adiabatically deforming an ensemble intriguing non-thermal states can be prepared, which have the potential to harbor interesting physics inaccessible by thermal pathways. For example we show that for Luttinger liquid non-thermal long ranged correlations can be stabilized or that in the Kitaev chain the non-thermal distribution function after the deformation can greatly alter the critical temperature separating topological and trivial phases.

To illustrate the main idea, we concentrate on initial thermal states w.r.t. some Hamiltonian $H$ and slowly vary some system parameter(s) to a final Hamiltonian $H'$. The system is assumed to be sufficiently isolated  from its environment such that during this slow deformation of $H\to H'$ it can be considered as closed. In this context "slow" refers to the regime in which the Gell-Mann and Low theorem holds for all excited states.\cite{Gell51,Brouder08} Of course the applicability of the Gell-Mann and Low theorem has to be checked on a case by case basis.
The theorem states that the eigenstates of the initial Hamiltonian are mapped to the eigenstates of the final Hamiltonian under an adiabatic passage. To simplify the general discussion we assume a non-degenerate discrete spectrum of the family of Hamiltonians which describe the deformation $H\to H'$. In our explicit examples studied below we will relax this assumption and test the range of validity of the results presented explicitly. 

For the two Hamiltonians $H$ and $H'$ with $H\left| E_n\right \rangle=E_n\left| E_n\right \rangle $ and 
$H'\left| E'_n\right \rangle=E'_n\left| E'_n\right \rangle $
the adiabatic deformation of a thermal ensemble $\rho =\frac{1}{Z}\sum_n e^{- E_n/k_{\rm B}T} \left| E_n\right \rangle \left\langle E_n\right|$
leads to 
\begin{equation}
\rho   \stackrel {\rm {adia}}{\rightarrow}\rho^{\rm deform} =\frac{1}{Z}\sum_n e^{- E_n/k_{\rm B}T} \left| E'_n\right \rangle \left\langle E'_n\right|
\end{equation}
with $\rho$ being the density operator. The deformed ensemble appears similar to a thermal one, but with changed Boltzmann factors corresponding to the initial Hamiltonian. A priori the consequences of this non-thermal distribution, e.g. for observables, are unclear. It is however obvious that the two limits of vanishing and infinite temperature trivially deform into an ensemble with the same zero or infinite temperature, respectively. For finite non-zero temperature of the thermal ensemble before the passage on the other hand, tuning the Boltzmann factors (by clever choice of the initial Hamiltonian)  allows for engineering and tuning the properties of the state after the deformation. 

We emphasize that the procedure we propose helps considerably in this engineering process. To illustrate this let us consider the most simple case, where the family of Hamiltonians during the deformation $H\to H'$ are non-interacting. The eigenmodes of any non-interacting model are easy to determine. Now if we consider first a sudden quench out of, e.g., the ground state of $H$ engineering the properties of the final state after the quench is still difficult as it involves the knowledge of the overlap of the eigenstates before and after the quench. Of course one could simply solve this problem numerically varying the initial Hamiltonian $H$, but a straightforward, intuitive picture remains illusive.  On the contrary, for an adiabatically deformed ensemble predicting the final state is very simple. The mode occupancies remain constant and only the energies of the modes follow the adiabatic passage. This allows for example to engineer the following simple, but interesting example. Let us assume we want to engineer a state which, e.g., displays a jump of arbitrary height of the mode occupancy around the chemical potential in a non-interacting non-gapped system. We can easily solve this problem using nothing but general arguments by considering a non-interacting band insulator in thermal equilibrium with the chemical potential in the middle of the band gap. If for simplicity the dispersion is assumed to be symmetric, the occupancy of the modes of this band insulator will take continuous values from $C_1$ to $1/2+Z/2$ (for the part of the spectrum below the chemical potential) 
and from $1/2-Z/2$ to $C_2$ (for the part of the spectrum above the chemical potential), with some values $C_1$,$C_2$, $0\leq Z\leq 1$ that depend on the temperature of the system (and trivially $T\to 0$, $Z\to 1$ and $T\to\infty$, $Z\to 0$). Adiabatically closing the band gap deforms the eigenstates keeping the Boltzmann weights constant, thus in the final Hamiltonian the weights of the modes will still be distributed from $C_1$ to $1/2+Z/2$ and from $1/2-Z/2$ to $C_2$ (below and above the chemical potential, respectively), displaying a "quasi-particle"-like jump of height $Z$ (which can be tuned by the temperature before the adiabatic passage). This explicitly allows to design all possible values of $0\leq Z\leq1$. No thermal ensemble can be used to obtain such a state in a non-interacting problem. Performing a similar Gedanken experiment for the sudden quench to engineer such a non-thermal state is obscured by the difficulty of calculating the overlaps of final and initial states. Thus this constitutes one example where we have utilized adiabatic deformation to engineer a non-thermal state with the desired property (jump at the chemical potential of height $0<Z<1$) we asked for. This example will be analyzed more formally as the second example discussed below. 

Following this logic it is also clear why (effectively) non-interacting systems that follow a linear dispersion relation cannot be adiabatically deformed away from a thermal state, if the adiabatic deformation only changes the prefactor of the linear dispersion from $v \to v'$, as in the example of a Luttinger liquid with time dependent interactions mentioned above. The Boltzmann weights after the deformation $\sim e^{-\beta \sum_k v k}$ can trivially be rewritten as   $e^{-\beta' \sum_k v' k}$ with changed inverse temperature $\beta'=\beta v/v'$.

Identifying and engineering intriguing non-thermal physics via the adiabatically deformed ensemble proposed above calls for case studies. One can argue that the assumption of adiabatic deformation of all eigenstates, severely limits the speed with which the passage has to be performed in the absence of energy gaps. We will explicitly test this in the examples studied below and find that even in the absence of gaps the physics of the deformed system behave adiabatic on energy (real-space) scales much larger (much smaller) than a scale set by the inverse speed (we will concentrate on linear ramps with speed $v$ in the following) of the deformation.\cite{Dora11}. Since the general idea of engineering non-thermal states by adiabatic passage  applies to arbitrary physical systems, we will put it to the test in a variety of different important models in the following, because  we believe that this strategy is necessary to access the generality of the proposed concept. Every model study is self-contained and can be accessed independently of each other. We thus structure the rest of the paper in sections discussing the separate  examples: section I deals with a simple single oscillator, section II with a dimerized chain of interacting spinless fermions, section III with a BCS superconductor and section IV with the topological Kitaev chain. Finally in section V we give a concluding summary.

\section{Oscillator}
First, we consider a single oscillator as a simple example. The oscillator is described by a harmonic as well as an anharmonic ($\sim x^4$) contribution
\begin{equation}
H=\omega_0\left(b^\dagger b +\frac{1}{2}\right)+\frac{g(t)}{4\omega_0^2}\left(b^\dagger + b\right)^4
\end{equation} 
The single oscillator is  initially in a thermal state with temperature $T$ and we set $\omega_0=1$. The anharmonicity $g(t)\geq 0$ is tuned from its initial value $g_{\rm ini}$ to $0$. For this simple (toy-)model it is easy to show that the dynamics remains adiabatic as long as the speed of the ramp $v\ll1$, as the level spacing $\Delta E \geq 1$.

\begin{figure}[t]
\centering
\includegraphics[width=\columnwidth]{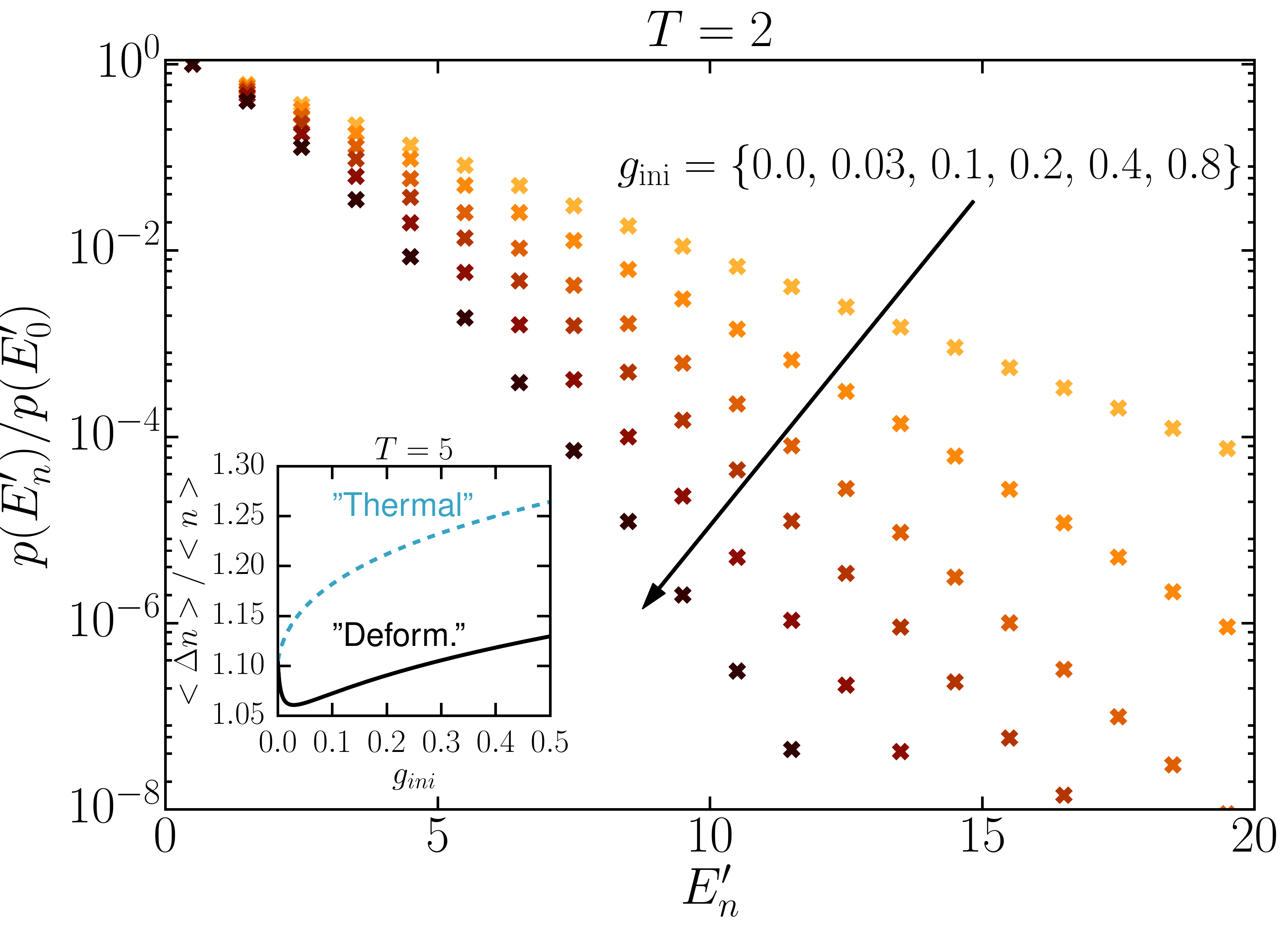}.
\caption{Main panel: adiabatic deformation of a single oscillator. We show the relative weights $p(E_n')/p(E_0')$ of the eigenstates $\left |E_n'\right \rangle$ to the ensemble after the adiabatic deformation from a thermal state with $T=2$. For $g_{\rm ini}>0$ the ensemble is clearly non-thermal as the  weights of the eigenstates $\left |E_n'\right \rangle$ fall off quicker than exponentially in their respective energies $E_n'$. The deformed ensemble thus suppresses  fluctuations into higher energy states. Inset: fluctuations in the mean boson occupancy $\Delta n=\sqrt{\left\langle n^2\right\rangle-\left\langle n\right\rangle ^2}$ relative to the mean boson occupancy $\left\langle n\right\rangle=\left\langle b^\dagger b\right\rangle$ itself after the adiabatic deformation from a thermal ensemble with $T=5$ (solid black line) as well as for a thermal state with temperature fixed by the mean occupancy after the deformation (dashed blue line).  }
\label{fig:Osc}
\end{figure}

Exemplary results of such an adiabatic deformation are summarized in Fig.~\ref{fig:Osc}. We find that the ensemble after the adiabatic deformation shows a clear non-thermal signature: the weight $p(E_n')$ of the eigenstates $\left |E_n'\right \rangle$ to the ensemble 
\begin{equation}
\rho^{\rm deform}=\frac{1}{Z}\sum_n p(E_n')\left |E_n'\right \rangle \left\langle E_n'\right|
\end{equation}
reached after the adiabatic deformation does not follow an exponential form $\sim e^{-E_n'/k_{\rm B}T}$ in their respective Eigenenergy $E_n'$ as would be the case in a thermal ensemble. In fact they fall off quicker then exponentially, indicating a suppression of fluctuations into higher occupation number states. In the inset we show the fluctuations in the mean boson occupancy $\Delta n=\sqrt{\left\langle n^2\right\rangle-\left\langle n\right\rangle ^2}$ relative to the mean boson occupancy $\left\langle n\right\rangle=\left\langle b^\dagger b\right\rangle$ itself after the adiabatic deformation as a solid black line. This signal to noise measure can be enhanced compared to a reference thermal ensemble (dashed blue line in the inset), where the effective temperature ${T_{\rm eff}}$ is determined by the mean boson occupancy $\left\langle n\right\rangle_{\rho^{\rm deform}}\stackrel{!}{=}\left\langle n\right\rangle_{T_{\rm eff}} = 1/(e^{1/{T_{\rm eff}}}-1)$. Thus $\Delta n$ highlights the highly non-thermal nature of the deformed ensemble.  We thus note that adiabatic passage can be used to suppress fluctuations beyond the cooling paradigm.

\section{Dimerized Chain}
Next we consider the 1D Hamiltonian 
\begin{equation}
H=\sum_{j=1}^{N-1} \left[J \left(c^\dagger_jc_{j+1}+{\rm {H.c.}}\right)+Un_{j}n_{j+1}\right]+\sum_{j=1}^{N}(-1)^j\frac{\delta(t)}{2}n_j
\label{eq:Hdimer}
\end{equation}
describing a dimerized nearest-neighbor tight-binding chain with open boundary conditions and density-density interaction $U$ between adjacent fermions (density $n_j=c^\dagger_j c_j$) and set $J=\hbar=k_{\rm B}=1$. We concentrate on half-filling. This model falls into the Luttinger liquid universality class at non-zero $|U|<2$ and $\delta=0$.\cite{Solyom79,Haldane80,Giamarchi03,Schoenhammer05}

{\it{Adiabatically Deformed Ensemble: $U=0$}} --- We discuss the non-interacting $U=0$ case first. This limit allows for a simple and exact treatment and reveals essential insights into the underlying physics of the problem. We calculate the adiabatic deformation of a thermal ensemble with $T=0.2$ starting from $\delta=1$ to $\delta=0$ for different deformation speeds $v$.  
\begin{figure}[t]
\centering
\includegraphics[width=\columnwidth]{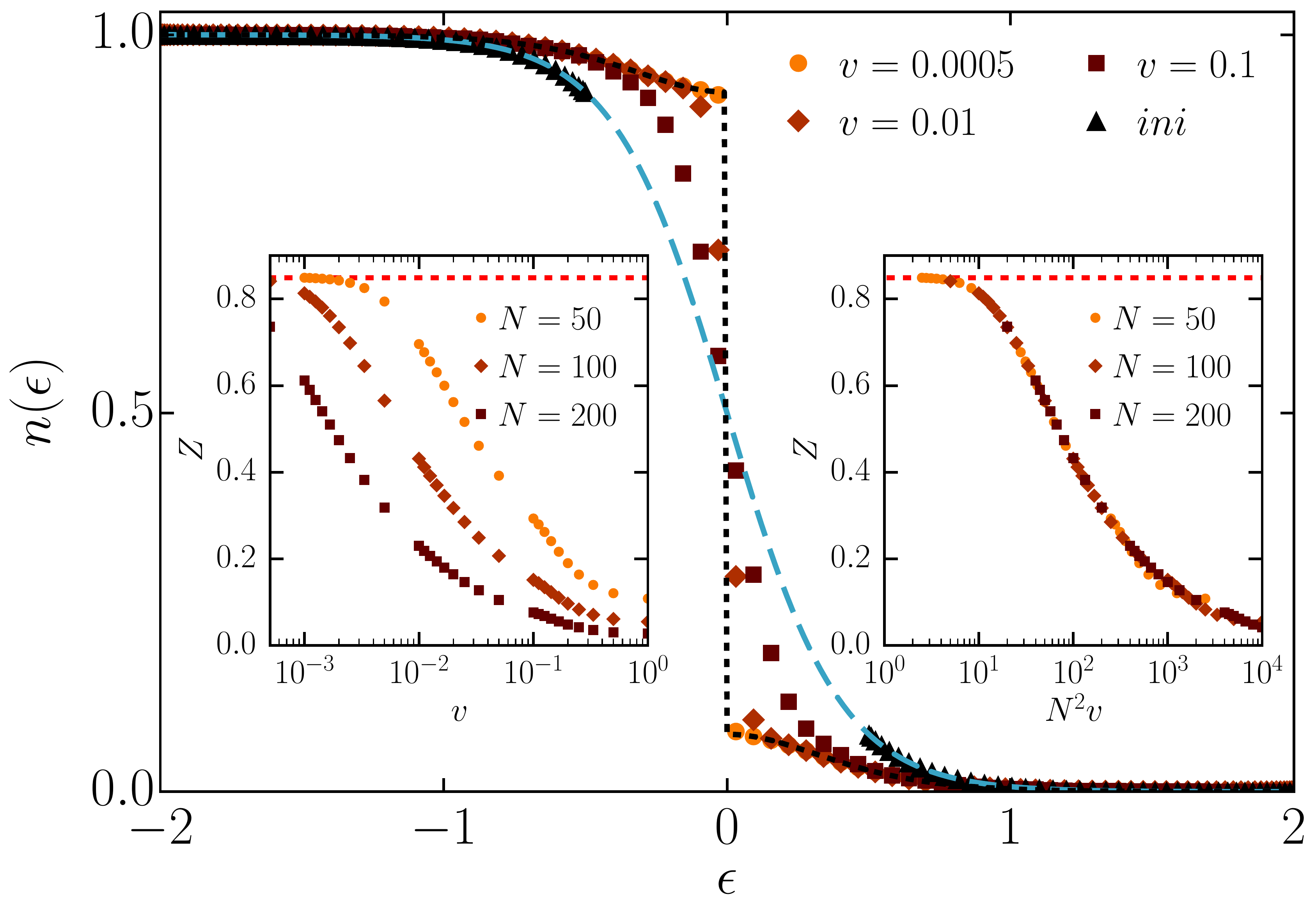}
\caption{Main panel: We deform the dimerized chain Eq.~\eqref{eq:Hdimer} from $\delta=1$ to $\delta=0$ and $U=0$. Shown is the distribution function before (black symbols) and after (colored symbols) the adiabatic passage for $N=100$. We initially prepare a thermal distribution with $T=0.2$ (thick dashed blue line). The thin dashed black line is the prediction Eq.~\eqref{eq;dimerpred}. The insets demonstrate the system size dependency: (left) as $N$ is increased slower time variations are needed to stay adiabatic, (right) all curves collapse when rescaling the x-axis by the minimum in single-particle energy gaps $\sim\!N^{-2}$. }
\label{fig:Firsttest}
\end{figure}
The results are summarized in Fig.~\ref{fig:Firsttest}. The main panel shows the distribution function $n$ in dependency of the energy $\epsilon$ (we suppress the momentum index $k$ in the energies $\epsilon_k$ in the following)  both initially (thermal Fermi-Dirac distribution (thick dashed blue line in Fig.~\ref{fig:Firsttest}), note the energy gap) as well as after the adiabatic deformation for different speeds $v$ and for $N=100$. For slow enough deformation  a  jump in the mode occupancy of finite height $Z=\lim\limits_{\epsilon\to 0} n(-\epsilon)-n(\epsilon)$ occurs around the Fermi edge $\epsilon=0$. This is in stark contrast to the thermal case at $T>0$. We derive the distribution function analytically (thin dashed black line in Fig.~\ref{fig:Firsttest})
\begin{equation}
n(\epsilon)=\frac{1}{e^{{\rm sign}(\epsilon)\sqrt{\epsilon^2+\delta(0)^2/4}}+1}.
\label{eq;dimerpred}
\end{equation}
It is clearly not a thermal one and one can engineer states exhibiting a very sharp jump at the Fermi edge of height 
\begin{equation}
c=\tanh\left(\frac{\delta(0)}{4T}\right).\label{eq:c} 
\end{equation}
resembling $T=0$ Fermi liquids.

Both insets demonstrate the dependency on system size. In the left inset we show that as the system size increases the energy gap between the states diminishes and therefore the speed $v$ for which the maximum $Z$ (dashed red line) is reached must be lowered. The right inset shows, that if the results are scaled w.r.t. the minimum in the single particle energy gaps $\Delta E \sim\!N^{-2}$, all curves collapse. Thus, we identify $v/\Delta E\ll 1$ as the  adiabatic limit.


%
%

{\it{Adiabatically Deformed Ensemble: finite $U$}}--- For the interacting model $U\neq 0$ we employ the DMRG, set up in the language of matrix product states, to describe its (thermo)dynamics.\cite{White92,Schollwock11,Mitra11,Karrasch12,Barther13,Kennes16,Mitra17} We concentrate on the same adiabatic ramp in $\delta$ as described above and formulate the DMRG  directly in the thermodynamic limit.\cite{Vidal07,Karrasch12a} An adiabatic evolution is strictly speaking not  possible in the infinite system (see above). However, we find that if the system is deformed with given speed, local observables supported on a spatial region of width $\sim 1/v$ still deform adiabatically.\cite{Dora11} In this looser sense locally the system appears adiabatic, where the degree of this locality can be tuned by the speed of the deformation. 

\begin{figure}[t]
\centering
\includegraphics[width=\columnwidth]{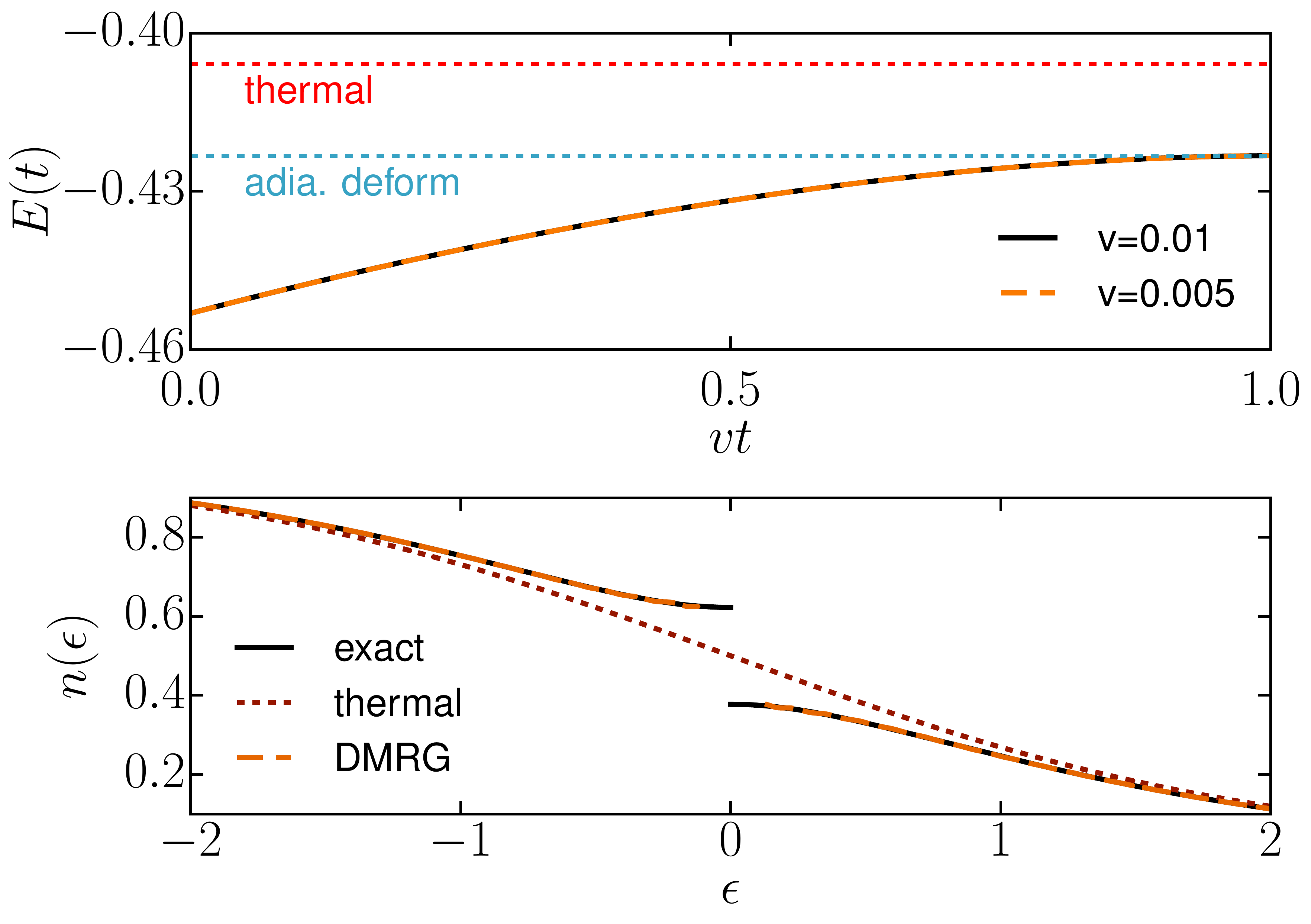}
\caption{ 
DMRG benchmark at $U=0$ for the same protocol as studied in Fig.~\ref{fig:Firsttest}. Upper panel: Energy per lattice site after an adiabatic deformation of $\delta=1$ to $\delta=0$ of Eq.~\eqref{eq:Hdimer} for
two deformation speeds and $T = 1$. We find good agreement with the expectation from an adiabatically deformed ensemble while there is a significant deviation from the thermal value at $T = 1$.  
Lower panel: Distribution function $n(\epsilon)$ after  the same adiabatic deformation as in the upper panel, with speed
$v = 0.01$ at $T = 1$ in the non-interacting limit $U = 0$ (dashed
orange line). We also show the prediction from Eq.~\eqref{eq;dimerpred} (solid black
line) as well as the thermal distribution (dashed red line).
}
\label{fig:DMRGtest_sub}
\end{figure}
As adiabatic deformation of initial states is not routinely studied in DMRG (for an exception see Ref.~\onlinecite{Schmitteckert13}) we first benchmark our DMRG result to those obtained analytically in the model of Eq.~\eqref{eq:Hdimer} at $U=0$. We concentrate on an adiabatic deformation of $\delta=1$ to $\delta=0$ as above. In all of our simulations we choose the parameters such that the results are converged on the scale of the plots. In practice this means that we employ a second order Trotter decomposition for the imaginary time evolution to prepare the initial thermal state of the system with a step size of $\Delta \beta=0.005$ at fixed bond dimension $\chi=300$. Subsequently, we subject this initial state to a real time evolution using a fourth order trotter scheme with step size $\Delta t=0.01$ and a dynamically increasing bond dimension, keeping the truncation error below a threshold of $10^{-8}$ at each time step. 

Fig.~\ref{fig:DMRGtest_sub} shows the benchmark comparing the DMRG results to the exact solution. 
The upper panel shows the energy per lattice site after an adiabatic deformation of $\delta=1$ to $\delta=0$ of Eq.~\eqref{eq:Hdimer} for two deformation speeds and $T=1$. This local observable behaves perfectly adiabatic at the speeds chosen although the system is infinite. We can predict the energy from Eq.~\eqref{eq;dimerpred} and find perfect agreement, while it deviates strongly from the thermal value at $T=1$.
The lower panels shows the distribution function $n(\epsilon)$ of the deformed ensemble ($v=0.01$) at $T=1$ in the non-interacting limit $U=0$ extracted from the Fourier transform of the correlator $c^\dagger_{L/2}c_{L/2+n}$. We restrict our  calculation to $|n|<100$. For the values of $\epsilon$ which can be extracted reliably with $|n|<100$ the agreement to the analytical result Eq.~\eqref{eq;dimerpred} is perfect.

\begin{figure}[t]
\centering
\includegraphics[width=\columnwidth]{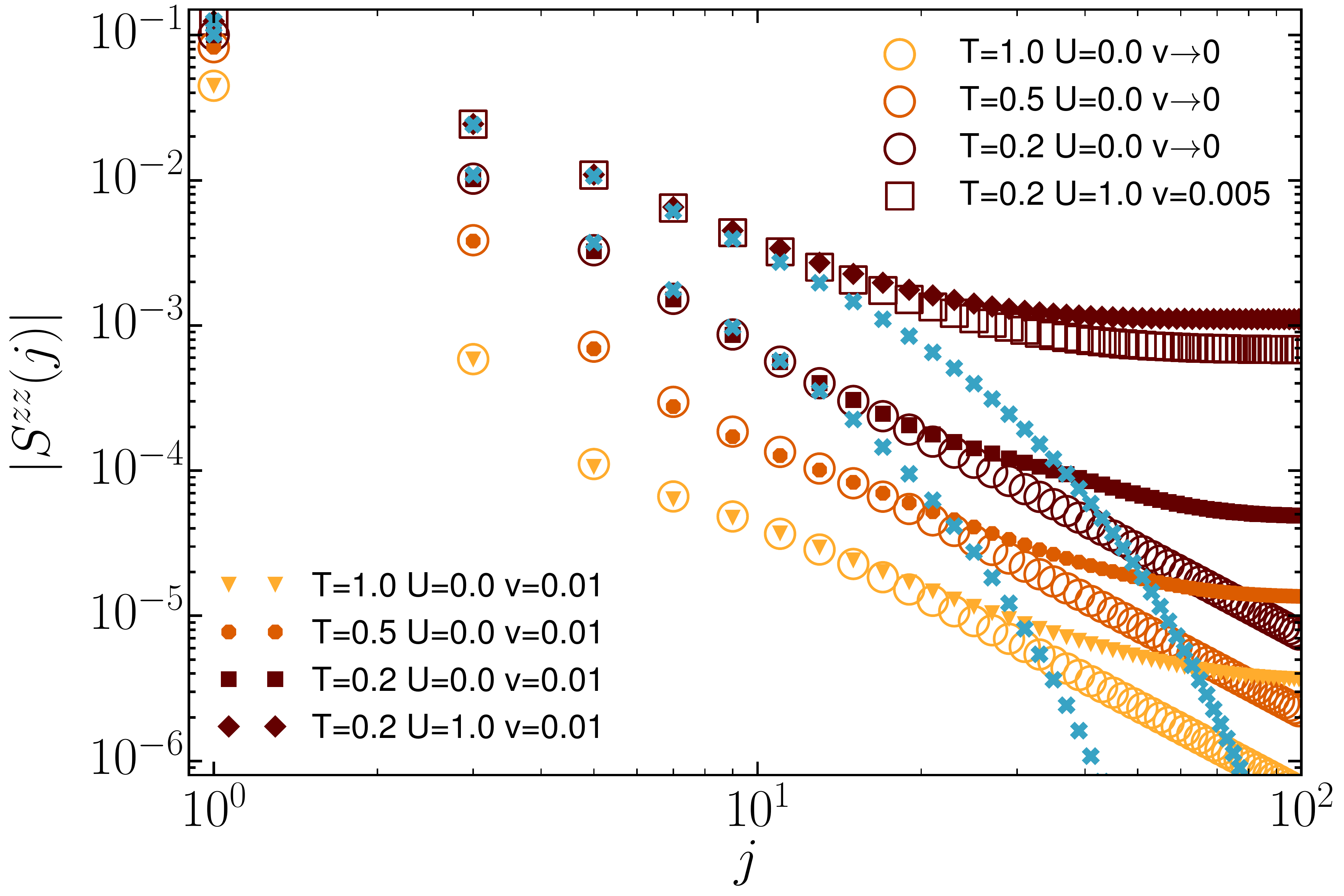}
\caption{ We study the same adiabatic deformation of $\delta$ of the dimerized chain Eq.~\eqref{eq:Hdimer} as in Fig.~\ref{fig:Firsttest}, but also showing non-zero $U$. We compare $v=0.01$ (smaller filled symbols) to $v\to 0$ in the non-interacting as well as a smaller $v=0.005$  in the interacting case (larger open symbols). For comparison we show the thermal expectation w.r.t. a conservatively low chosen temperature $T=1/16<T_{\rm eff}$, where $T_{\rm eff}$ is the effective temperature fixing the correct energy expectation value at the end of the deformation (blue crosses). In the adiabatically deformed ensemble the cut-off temperature is removed.  
}
\label{fig:DMRGtest}
\end{figure}

Our DMRG results beyond the consistency check of Fig.~\ref{fig:DMRGtest_sub} are summarized in Fig.~\ref{fig:DMRGtest}. We concentrate on the density-density correlator $S^{zz}(j)=\left\langle(n_{L/2}-1/2)(n_{L/2+j}-1/2)\right\rangle$. In thermal equilibrium  and for $\delta=0$
this function decays exponentially for distances $j$ larger than a
characteristic scale of the order of the inverse temperature $1/T$.\cite{Giamarchi03}  The correlator $S^{zz}(j)$ calculated for a thermal ensemble with  temperature $T=1/16$ is shown as the x'es in Fig.~\ref{fig:DMRGtest} for $U=0$ and $U=1$. We choose this value of the temperature as it is smaller than $T_{\rm eff}$, determined by $\left\langle H\right\rangle_{\rho^{\rm deform}}\stackrel{!}{=}\left\langle H\right\rangle_{T_{\rm eff}}$, for all shown parameter sets. The small filled symbols in Fig.~\ref{fig:DMRGtest} show $S^{zz}(j)$ obtained after a $\delta$  ramp from $\delta=1$ to $\delta=0$ with velocity $v = 0.01$. We want to study the degree to which these results are in the adiabatic limit and  thus compare these results to a slower deformation. In the non-interacting case we can compare to results obtained in the limit $v \to 0$ while in the interacting case we choose a twice smaller velocity $v = 0.005$ for comparison. The results obtained at this smaller velocity are given as open larger symbols in Fig.~\ref{fig:DMRGtest}. 
The results indicate that the more
adiabatic the deformation becomes the more the region extends in space over which the results for the two different deformation speeds agree.  For large $j$ the two curves start to deviate indicating that the adiabatic regime is left and to obtain converged results at these values of $j$  lower speeds need to be considered. Here we want to concentrate on the spatial regime, where the results are converged with respect to the speed of deformation $v$.

In equilibrium at $T=0$, $S^{zz}(j)$ displays long ranged correlations 
following a sum of two power-laws  in real space. This type of power-law decay composed of a momentum $q=0$
and a $q=2k_F$ component (with $k_F$ being the Fermi momentum) is
characteristic for Luttinger liquids.\cite{Voit94} For the adiabatically deformed ensemble we find that as we lower the speed $v$, the regime over which  long-ranged power-law correlations can be observed, similar to the $T=0$ case, extends. 
Thus by lowering $v$, as in the equilibrium case by lowering $T$, one can stabilize  long ranged correlations over an increasing spatial regime, where in the adiabatically deformed ensemble the speed $v$ has replaced the temperature $T$ as a cutoff in real space. By comparing to a temperature chosen smaller than the effective temperature fixed by the energy expectation value after the adiabatic deformation (illustrated by x'es in Fig.~\ref{fig:DMRGtest}), we see that we go beyond the cooling paradigm and extend the long ranged correlations beyond the thermal cutoff.   
Slow ramping thus  opens up a route towards an
experimental realization of Luttinger liquid behavior in systems which
cannot be cooled down sufficiently.

\begin{figure}[t]
\centering
\includegraphics[width=\columnwidth]{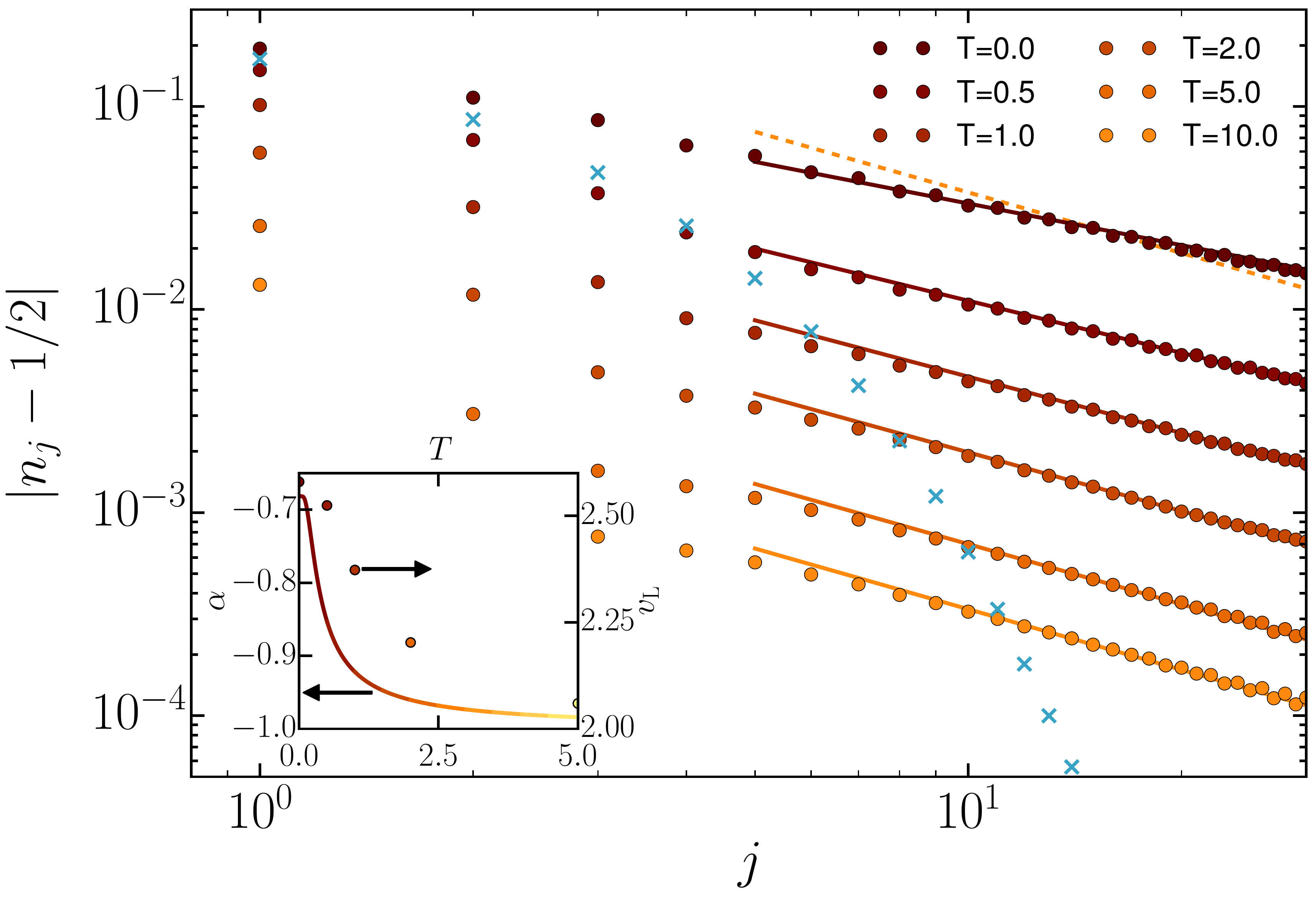}
\caption{Quasi-steady occupancy deviations from one half found at one of the boundaries of the system (dots) after an adiabatic deformation from $\delta=1$ to $\delta=0$ of Eq.~\eqref{eq:Hdimer} at $U=0$ and a subsequent interaction quench to $U=1$. The quasi steady state found shows a power law dependent behavior of the occupancy deviations, with an exponent Eq.~\eqref{eq:alphaB_red} (lines). To show that the exponent has changed sufficiently to distinguish different temperatures we add also the power-law of the largest temperature ($T=10$) to the dots of the smallest one ($T=0$) as a dashed line. The exponent $\alpha$ is temperature dependent and can be tuned from its maximal value at $T=0$ to the non-interacting limit at $T=\infty$. Blue crosses indicate the corresponding result found in thermal equilibrium using $T=0.25$. Clearly temperature in thermal equilibrium cuts of the power-law at site $j\sim 1/T$. This cutoff is removed by the adiabatic deformation, because the jump at the Fermi edge is not softened in the adiabatically deformed ensemble. The inset shows the analytic prediction of the exponent of the power-law (line) as well as the renormalization of the velocity $v_{\rm L}$. The velocity is tuned form its maximal value at $T=0$ to the non-interacting value at $T=\infty$ in resemblance to $\alpha$.     }
\label{fig:LL_qunech}
\end{figure}

{\it{Probing an Adiabatically Deformed Ensemble at $U=0$ with a Subsequent $U$ Quench}}--- To analyze the non-thermal behavior obtained by deforming the dimerized chain Eq.~\eqref{eq:Hdimer} further, we additionally perform a study of an interaction quench, by abruptly turning on the density-density type of interaction  
$$
H_U= \sum_{j=1}^{N-1} U n_jn_{j+1}
$$ 
{\it{after}} the adiabatic passage from $\delta=1$ to $\delta=0$ at $U=0$ has been completed ($\delta=0$ was reached). This is the same protocol as studied above, with the crucial difference being that the adiabatic deformation is done without interaction first and only subsequently the interaction is turned on abruptly. We use a functional renormalization group (FRG) approach as described in Ref.~\onlinecite{Kennes12,Kennes13} to address the long-time asymptotic and analyze power-law correlations unambiguously. Being able to address large system size and times comes at the cost of the method being approximate in the interaction strength. We concentrate on the occupancy deviations from half-filling. We find that after the interaction quench one wave  propagates from each end of the chain with velocity $v_{\rm L}$, leaving behind a distortion pattern in the occupancies. For times for which the wave of the one end of the  chain has passed a certain site, but before the wave of the other end reaches this particular site, a quasi-steady value of the occupancy is obtained. We will concentrate on these quasi-steady values of observables from now on and briefly review the results found for the quench in the thermal case first.\cite{Kennes13,Schiro14}
The FRG approach employed only captures the leading behavior in $U$ of the {\it{exponent}} of boundary or impurity physics. To this order the equilibrium and the quenched non-equilibrium exponent of boundary and impurity physics agree. Here, we concentrate on how the occupancies, exhibiting Friedel oscillations after the quench, fall off from one boundary of the chain. In the quenched case starting from a thermal $T=0$ ensemble (ground state) those follow $|n_j-1/2|\sim j^{-\alpha} $.   Expanding the expression for the exponent $\alpha=-(K^2+1)/2$ in Ref.~\onlinecite{Kennes13} to leading order we find (using the Bethe ansatz results for the Luttinger parameter\cite{Giamarchi03,Schoenhammer05} $K=\pi/[2\arccos(-U/2)]$)
\begin{equation}
\alpha\approx -1+\frac{U}{\pi}.
\label{eq:alphaB_full}
\end{equation}
The interaction dependent (critical) power-law decay of the Friedel oscillations, being a hallmark of Luttinger liquid physics, is a consequence of the jump of the distribution function at the Fermi edge.\cite{Andergassen04,Enss05,Kennes13} This can most easily be identified by considering the flow equations in equilibrium which show that the distribution $n(\epsilon)$ and the prefactor $U$ of the flow equation show up only as a product $U [1-2n(\epsilon)]$. As a consequence reducing the height of the jump of the Fermi function by a factor of $c<1$ is equivalent to reducing $U$ by the same factor when focusing on low energy scales.   
Thus, in the case where the height of the jump at the Fermi edge is reduced below $1$ to $c$, we instead of Eq.~\eqref{eq:alphaB_full} expect
\begin{equation}
\alpha\approx -1+c\frac{U}{\pi}.
\label{eq:alphaB_red}
\end{equation}

The numerical results of our FRG study (colored dots) are summarized in Fig.~\ref{fig:LL_qunech} for different initial temperatures (of the ensemble before adiabatic deformation). We first perform an adiabatic deformation from $\delta=1$ to $\delta=0$ of Eq.~\eqref{eq:Hdimer} at speed $v=0.0005$ and $N=200$ and then abruptly turn on $U=1$. We depict the quasi-steady occupancies $n_j$. For reference the same quench is performed out of a thermal equilibrium ensemble w.r.t. $T=0.25$ (blue crosses). Clearly the cut-off in the power law behavior present in the quench out of an initial thermal ensemble is removed in the quench starting from the adiabatically deformed ensemble and the power-law behavior can be observed over sizable longer distance even at the highest initial temperature of $T=10$. The prediction Eq.~\eqref{eq:alphaB_red} with $c$ given by Eq.~\eqref{eq:c} (lines) is in good agreement with the numerical results. The adiabatically deformed ensemble can be used to tune the boundary exponent of the power-law decay of the occupancies from its maximal value $\alpha\approx-0.7$  at initial $T=0$ (same as for quench starting from thermal $T=0$ ensemble) all the way to the non-interacting limit of $\alpha=-1$ reached for $T=\infty$ (where the adiabatically deformed ensemble remains a trivial thermal ensemble with $T\to\infty$). This dependency of the boundary exponent is shown in the inset, which for completeness also shows the numerically extracted velocity $v_{\rm L}$, the second parameter characterizing a Luttinger liquid completely. $v_{\rm L}$ is tuned w.r.t. the temperature within the same limits (at $T=0$, $v_{\rm L}\approx 2.6$ (same as for quench starting from thermal $T=0$ ensemble) and at $T\to \infty$, $v_{\rm L}=2$, which is the non-interacting value).

To sum up, also in the case of first deforming the ensemble with respect to $\delta$ at $U=0$ and subsequently tuning on the interaction, we can stabilize a (critical) power-law behavior in observables (in this case how Friedel oscillations fall off away from a boundary), over a spatial region much larger than in the corresponding thermal case. Temperature is effectively removed as a cutoff and replaced by the inverse of the system size or the speed of deformation, whichever is larger.

{
\begin{figure*}[t]
\centering
\includegraphics[width=.48\columnwidth]{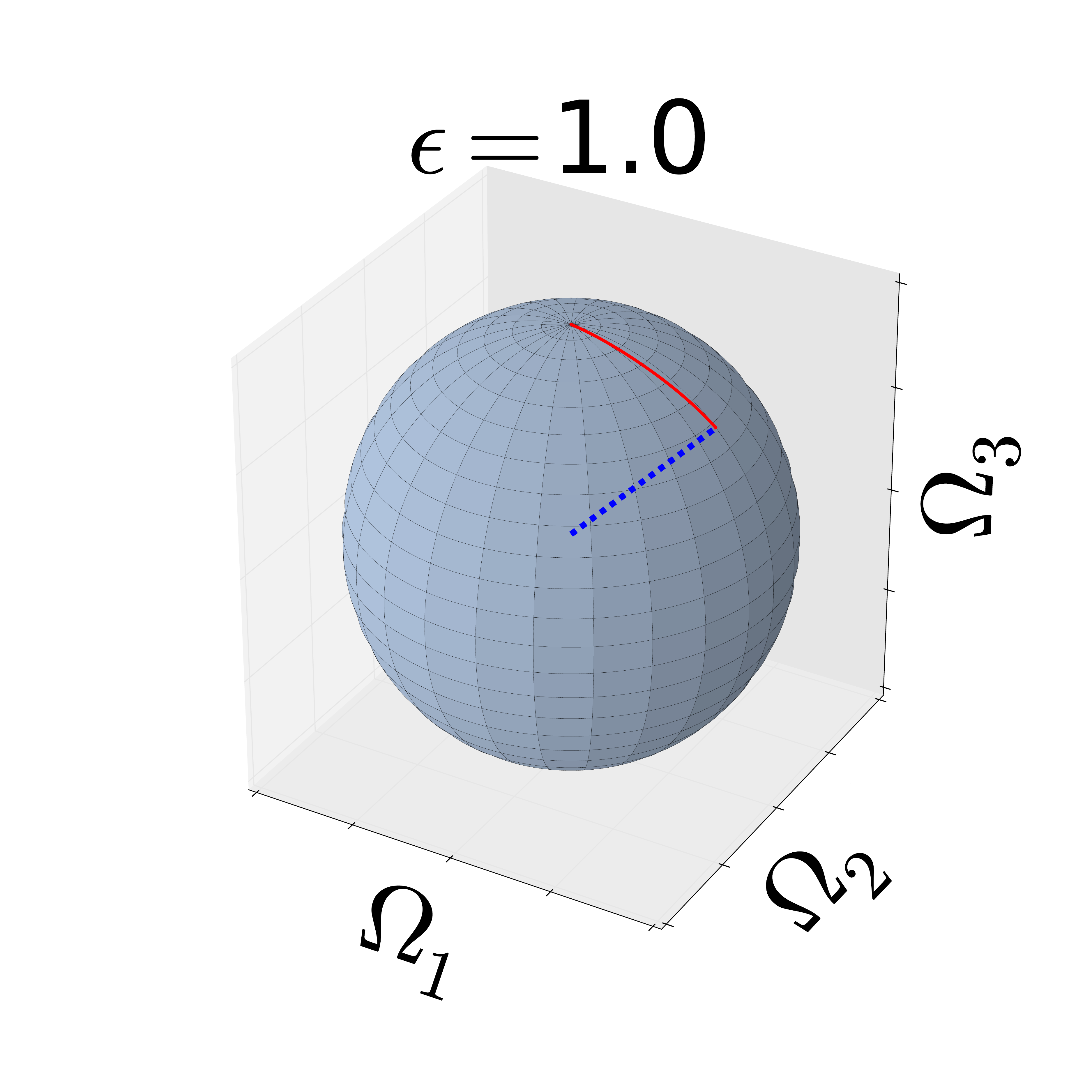}
\includegraphics[width=.48\columnwidth]{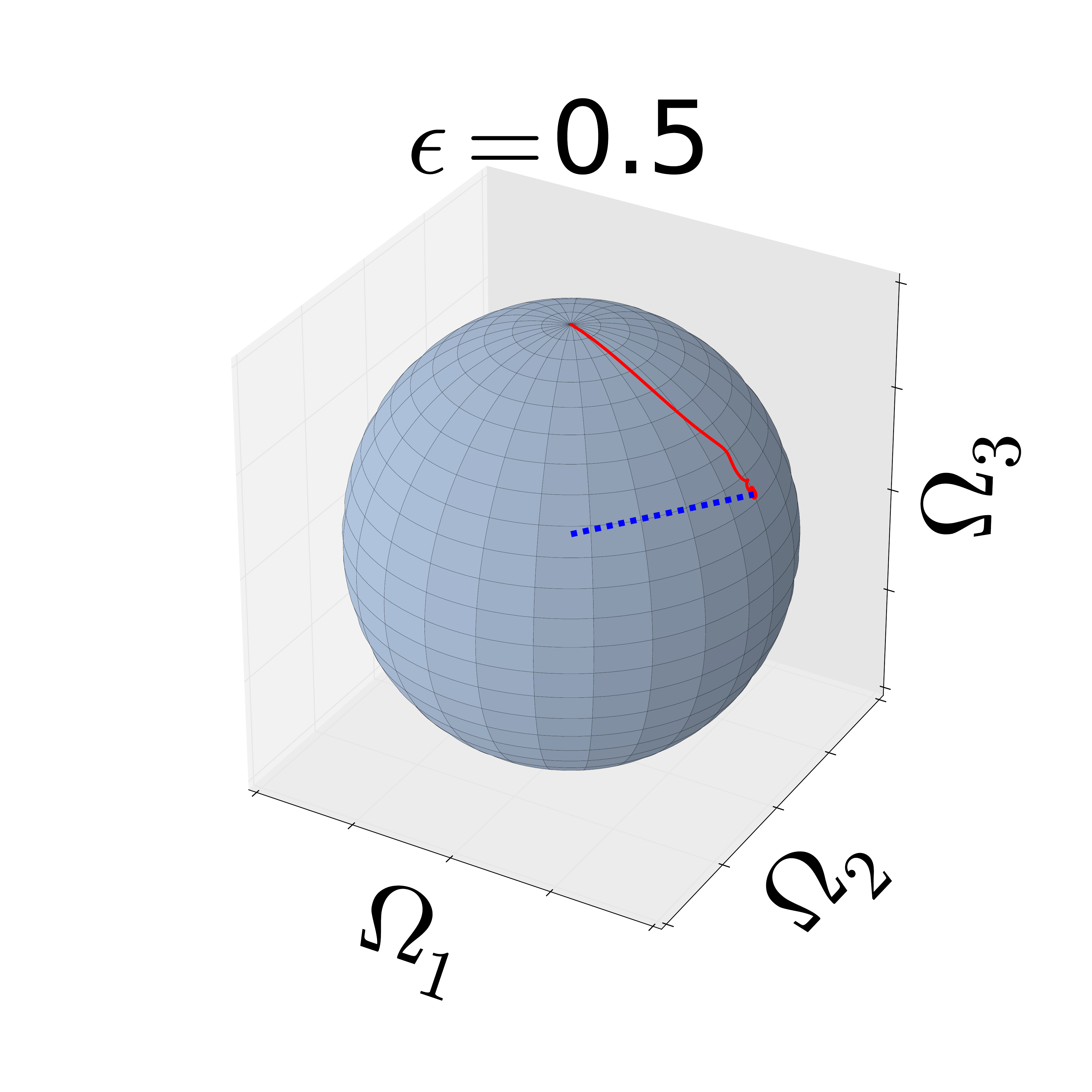}
\includegraphics[width=.48\columnwidth]{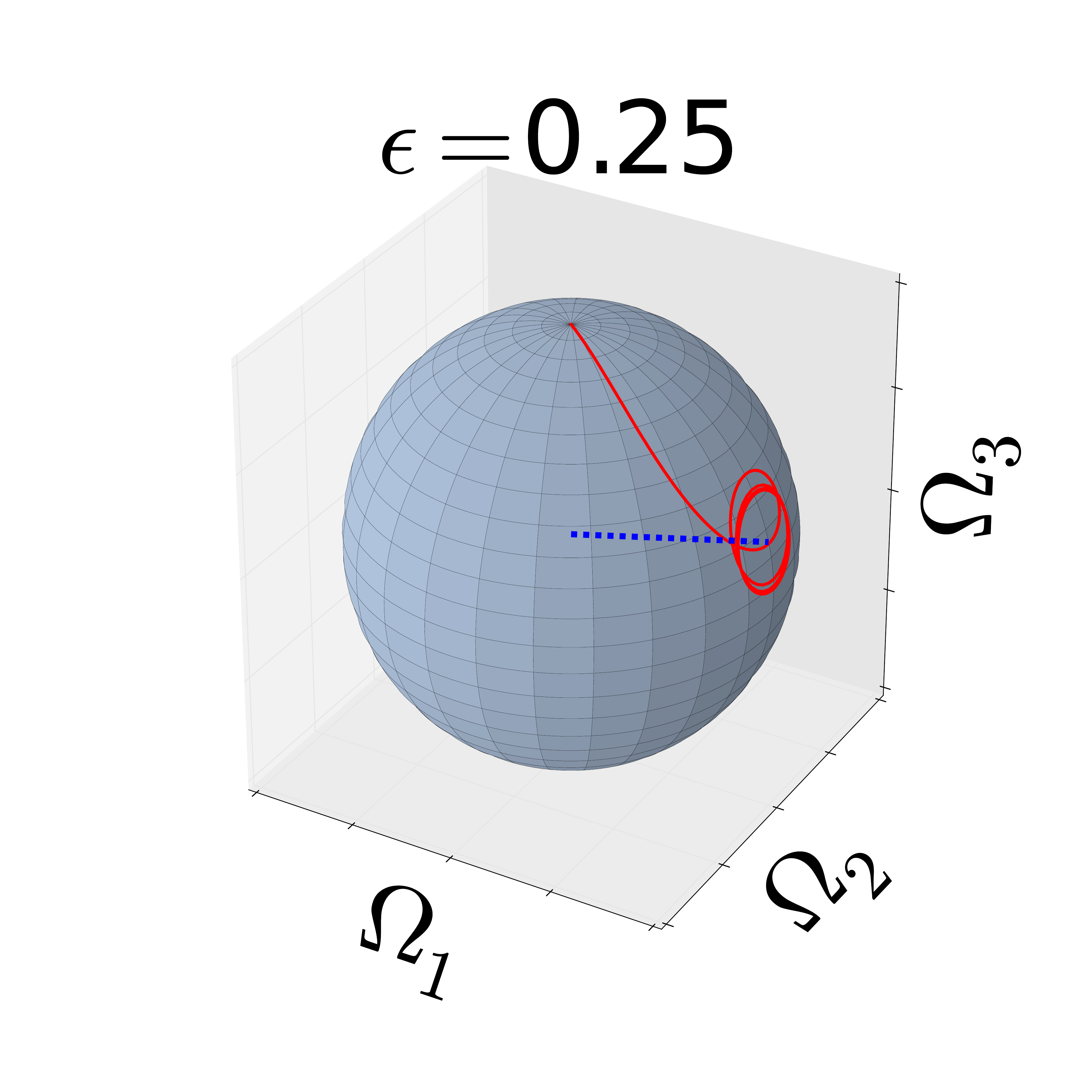}
\includegraphics[width=.48\columnwidth]{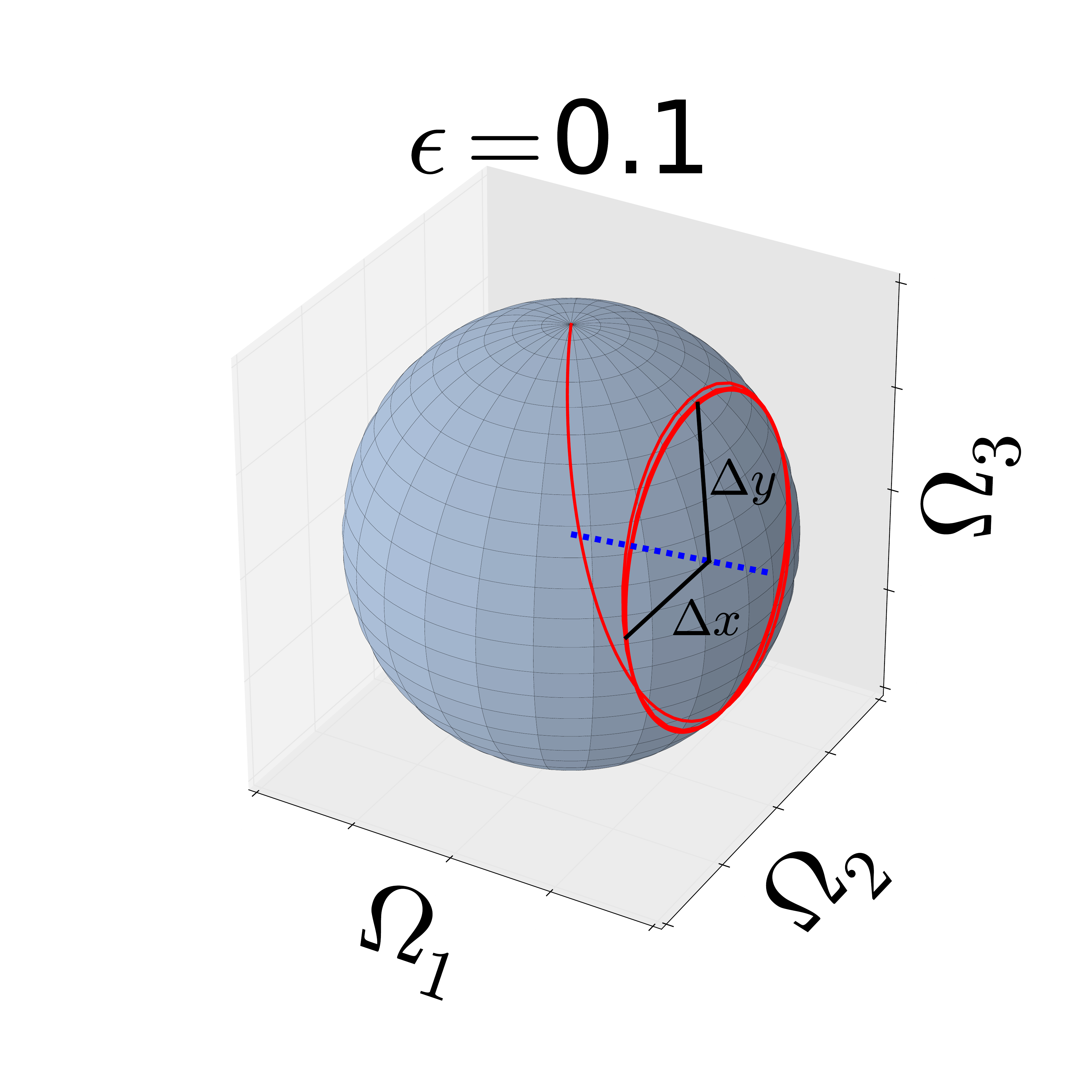}
\includegraphics[width=.48\columnwidth]{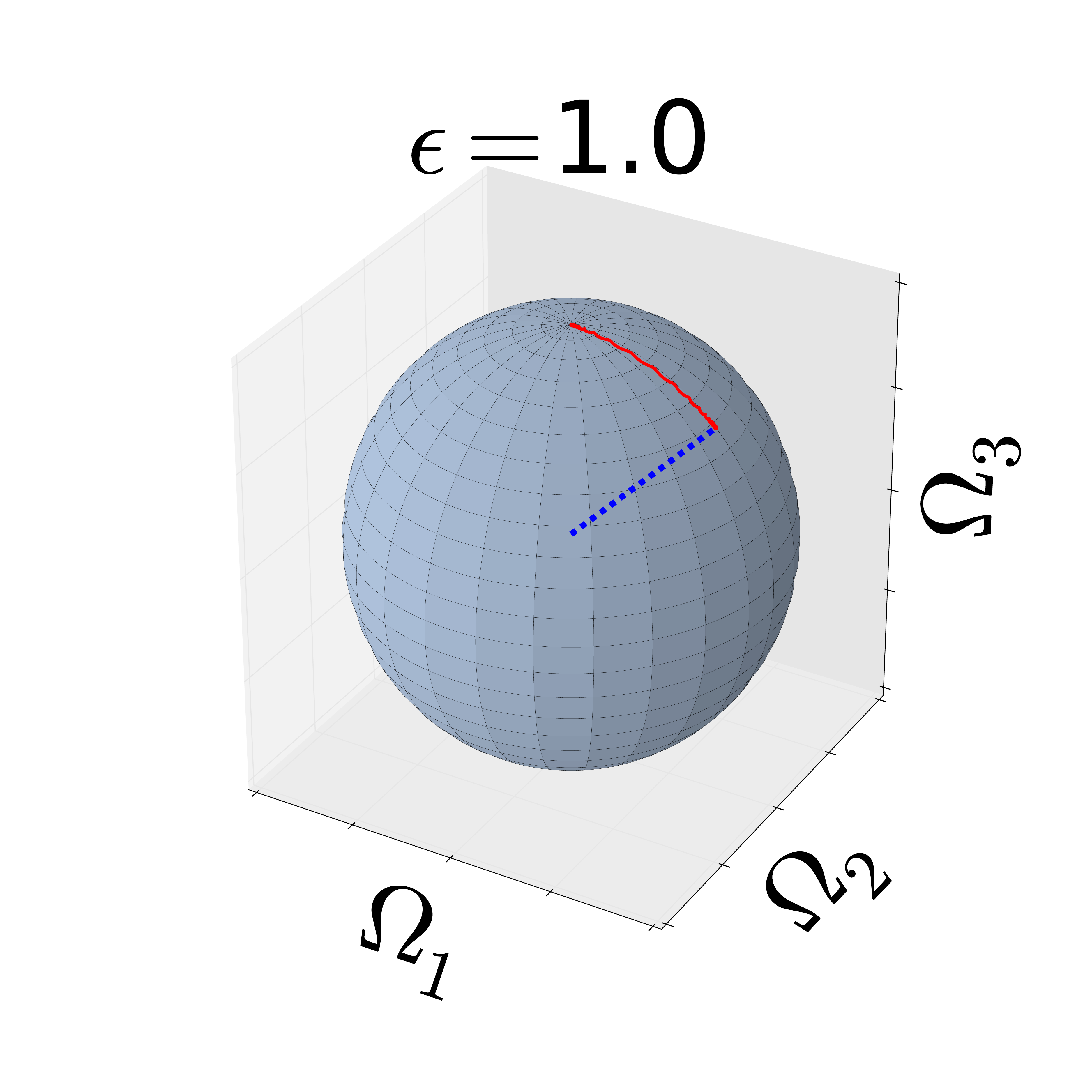}
\includegraphics[width=.48\columnwidth]{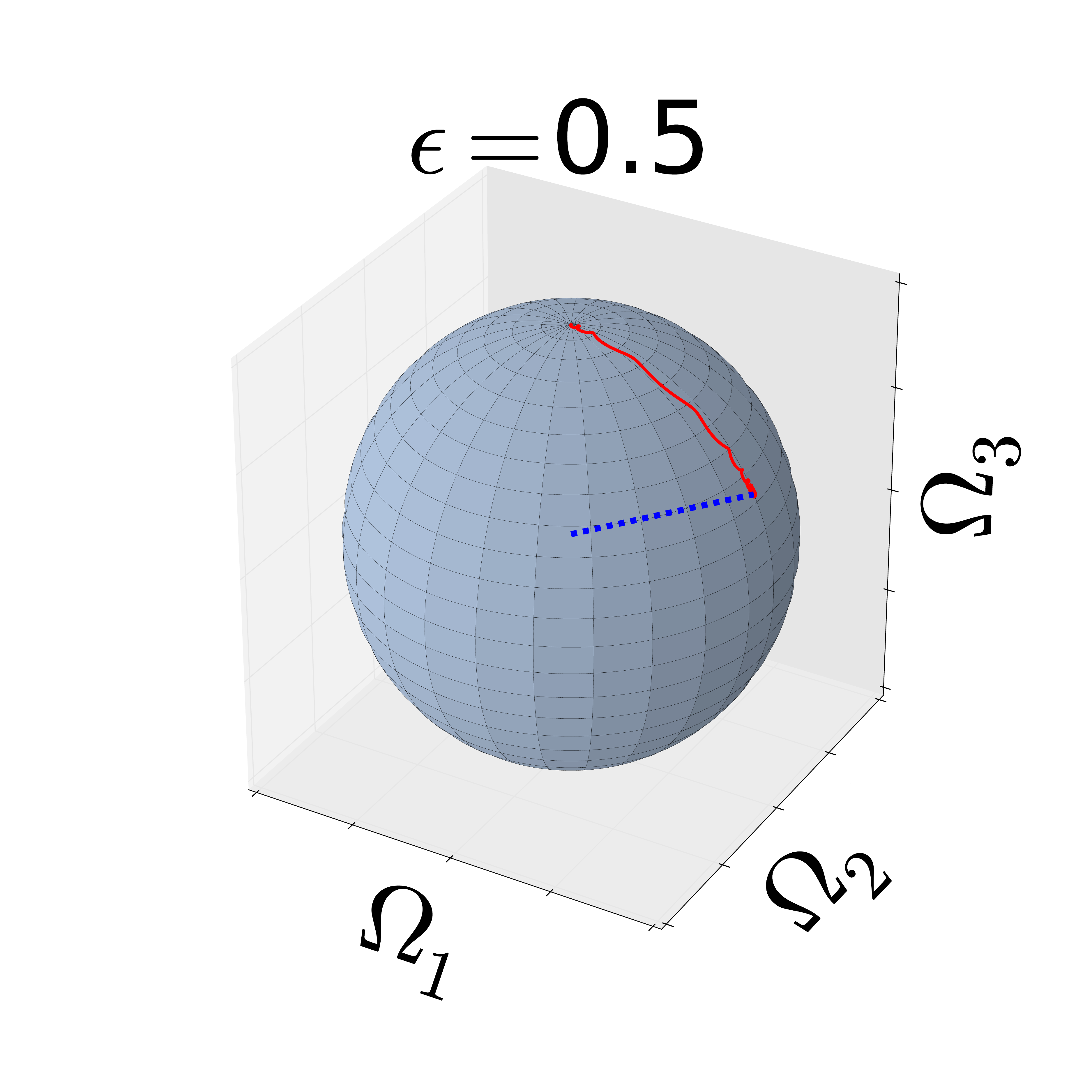}
\includegraphics[width=.48\columnwidth]{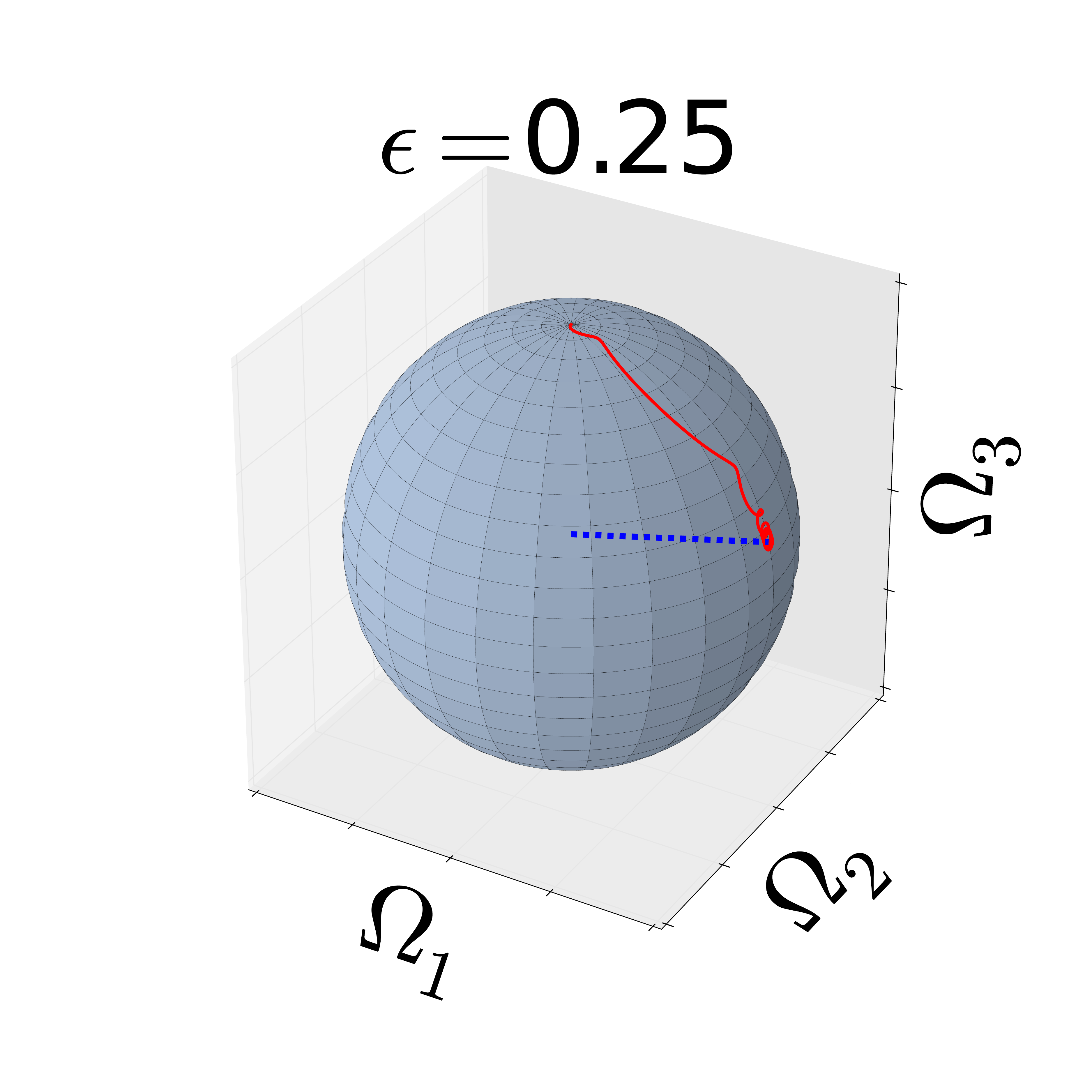}
\includegraphics[width=.48\columnwidth]{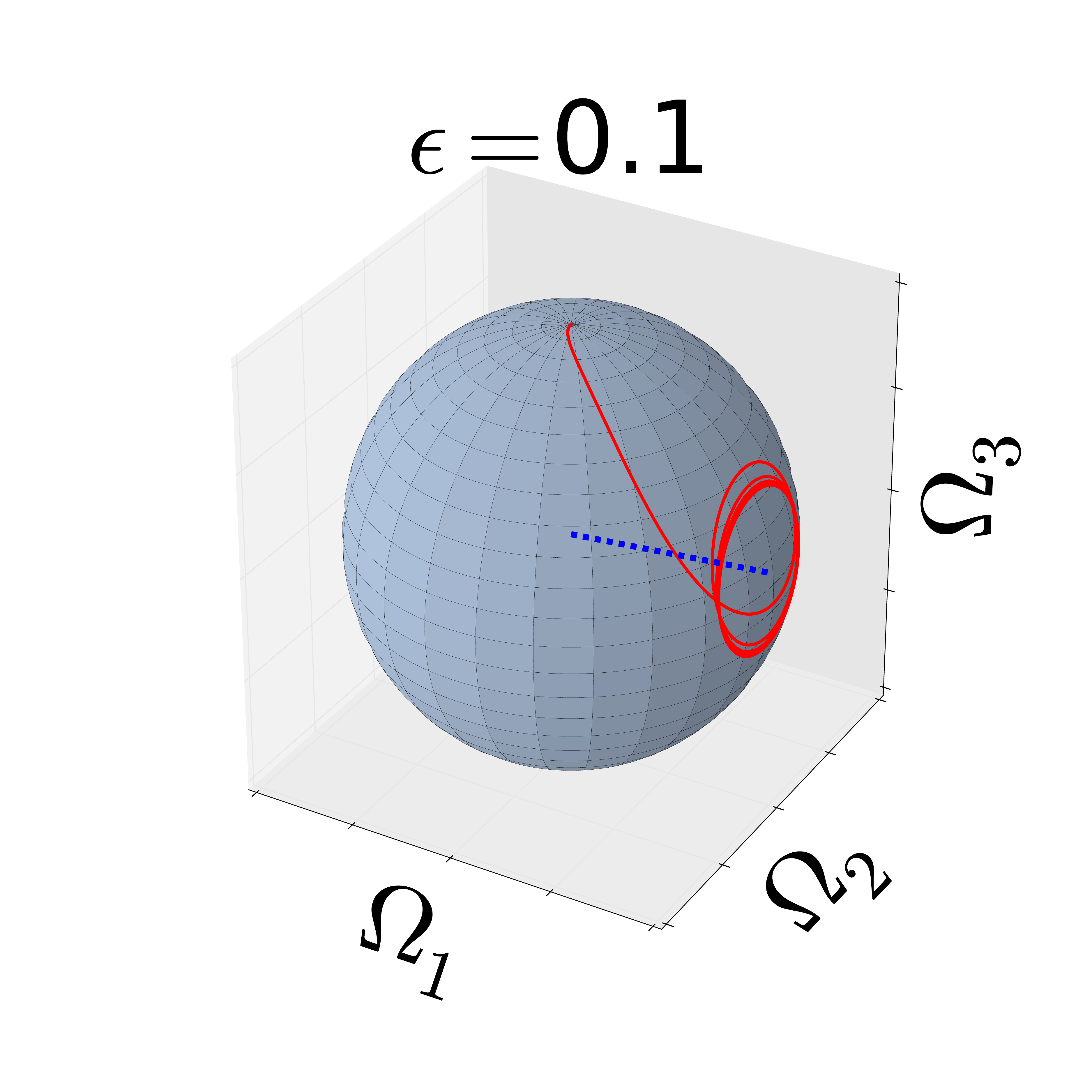}
\includegraphics[width=.96\columnwidth]{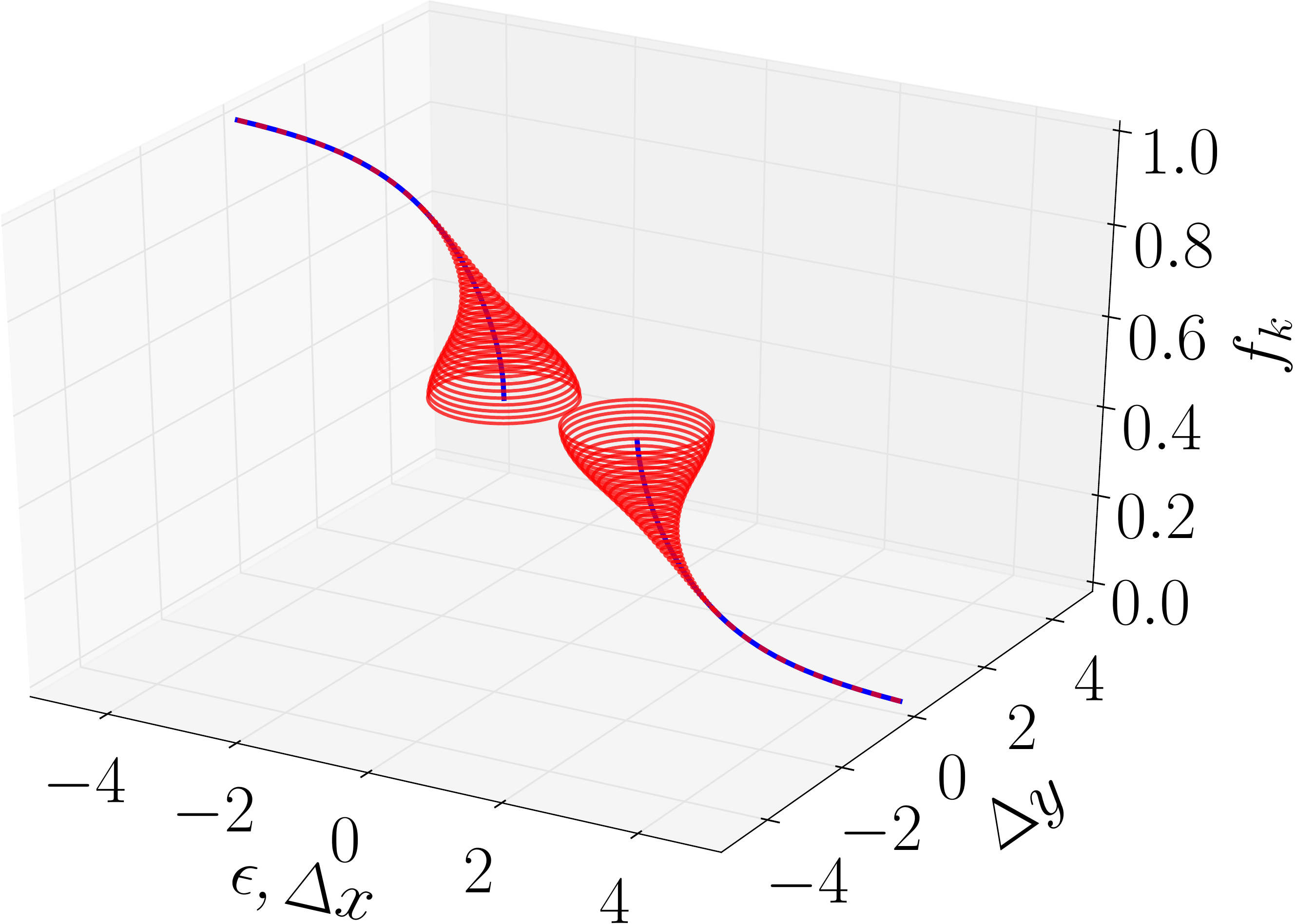}
\includegraphics[width=.96\columnwidth]{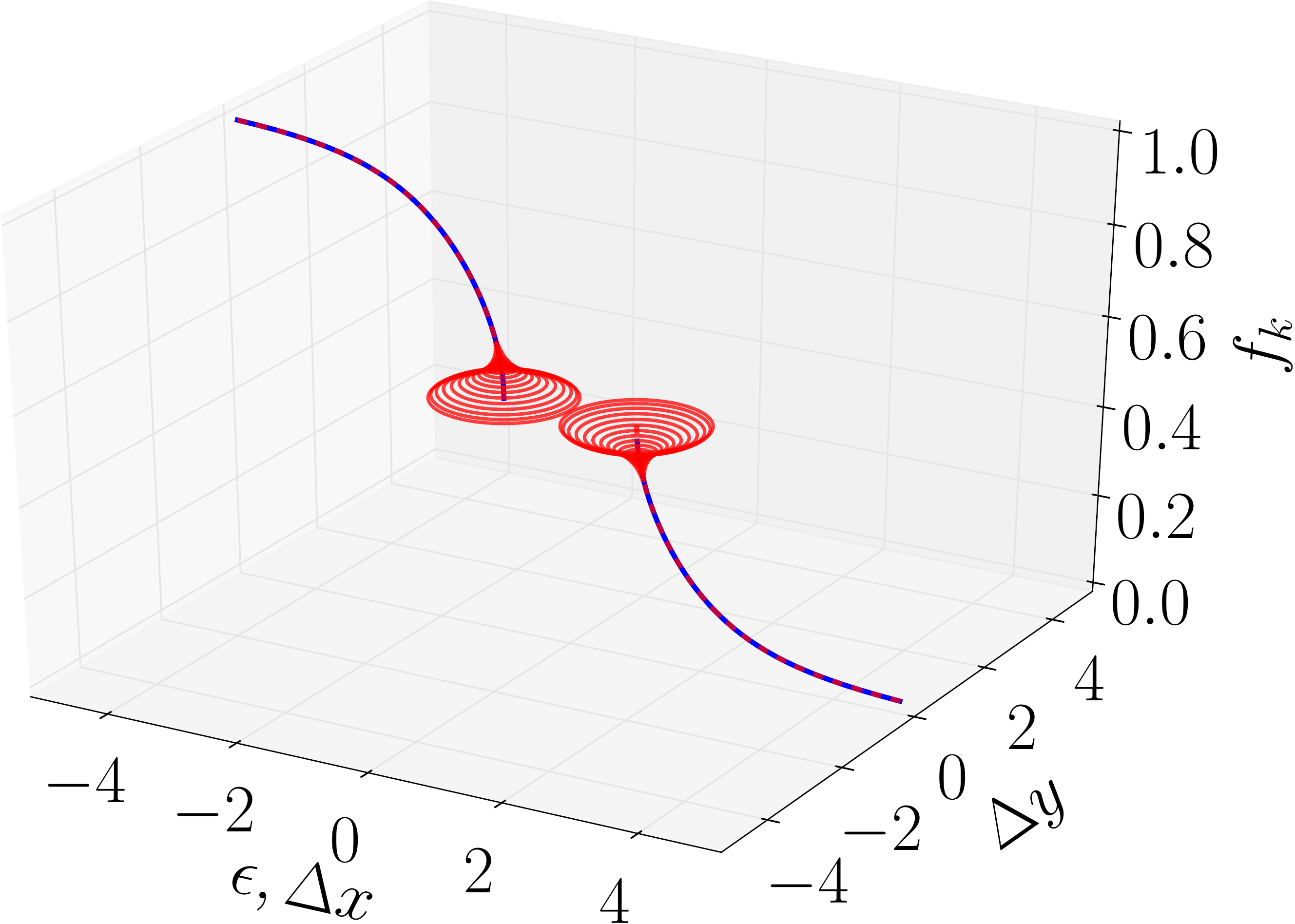}
\caption{Top two rows: Time evolution of $\vec \Omega$ (red line) for different $\epsilon$ and two $T_0=5$ (upper row) and $T_0=10$ (lower row) on the Bloch sphere and $T=1$. Blue dashed lines indicate the adiabatically deformed prediction. Large $\epsilon$ or $T_0$ clearly lead to an adiabatic deformation in accordance with Eq.~\eqref{eq:BCS_ad}. The far right plot in the upper row includes the definition of $\Delta x$ and $\Delta y$ (black lines) used in the bottom row plots. Bottom row: Adiabatically deformed ensemble prediction (blue line) compared to full numerics (red line) for $T_0=1$ (left) and $T_0=5$ (right) for $T=1$. Red circles indicate the magnitude of deviation (defined as in the upper right plot) from the adiabatically deformed ensemble which lie in the plane perpendicular to the adiabatically deformed $\vec{\Omega}$. Clearly deviations vanish for $1/\epsilon/T_0\ll 1$. }
\label{fig:Blochs}
\end{figure*}

}

\section{BCS Superconductor}
 \begin{figure*}[t]
\centering
\includegraphics[width=2\columnwidth]{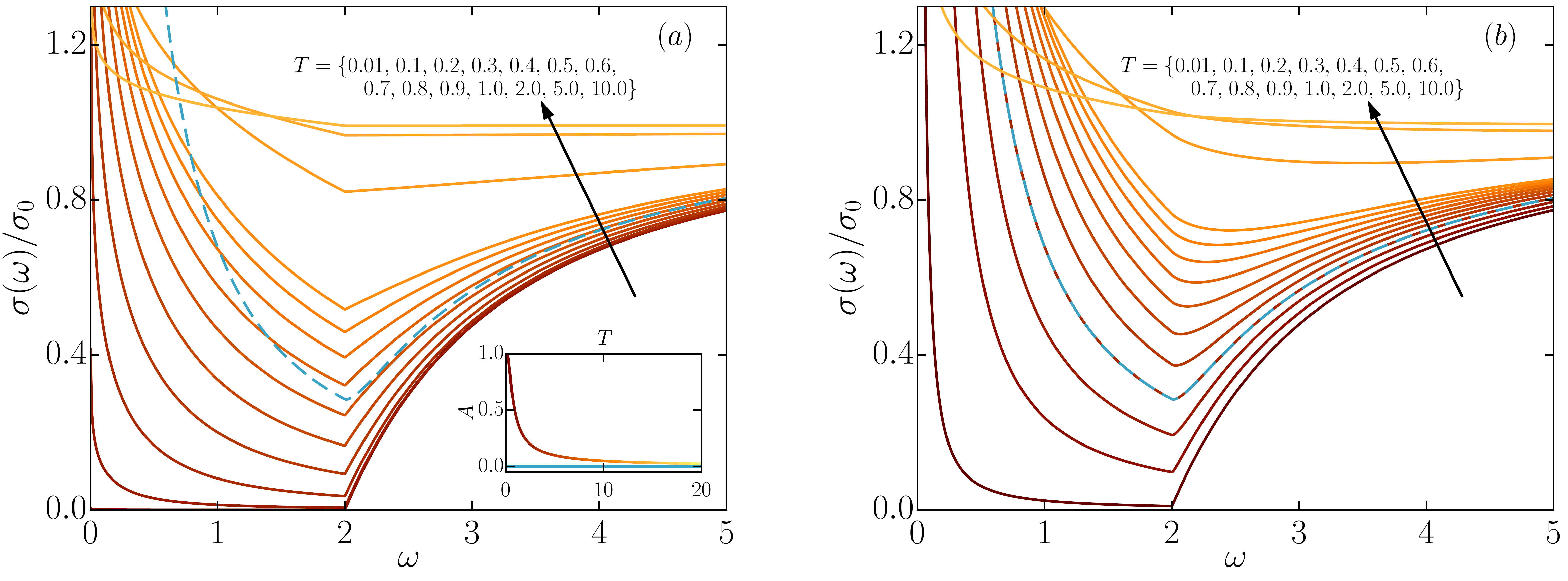}
\caption{Optical conductivity (black to yellow solid lines) for the thermal case (a) as well as the adiabatically deformed ensemble (b). For comparison the optical conductivity of the adiabatically deformed ensemble at initial temperature $T=0.3$ is included in both panels (blue dashed line). The inset shows the superfluid fraction $A$ for the thermal (multi-colored line) as well as the adiabatically deformed ensemble (blue line). The latter is zero.  }
\label{fig:noneq_cond}
\end{figure*}

As another example we study a BCS superconductor
\begin{equation}
H=\sum\limits_{k,\sigma=\uparrow\downarrow} \epsilon_k c_{k,\sigma}^\dagger c_{k,\sigma} + \Delta(t) \sum\limits_k  c^\dagger_{k,\uparrow} c^\dagger_{-k,\downarrow}+{\rm H.c.},\label{eq:Hini}
\end{equation}
with a time dependent gap $\Delta(t)$. Here  $c^{(\dagger)}_{k,\sigma}$ annihilates (creates) a fermion in the single particle state characterized by momentum $k$ and spin $\sigma$. We employ the language of Nambu-vectors $\Psi_k^\dagger=(c^\dagger_{k,\uparrow},c_{-k,\downarrow})$ to rewrite the Hamiltonian as  
\begin{equation}
H=\sum\limits_{k} \Psi^\dagger_k \begin{pmatrix}
\epsilon_k&\Delta(t)\\
\Delta(t)^*&-\epsilon_k
\end{pmatrix}\Psi_k.
\label{eq:Hsc}
\end{equation}

First, we concentrate on the case of a given gap function 
\begin{equation}
\Delta(t)=\frac{\Delta_0}{2}\left[\tanh\left(\frac{t-\delta}{T_0}\right)+1\right],
\end{equation}
where we slowly tune from a normal conducting to a superconducting state.
The unit of energy is set by $\Delta_0=1$. 

We exploit the representation of Ref.~\onlinecite{Barankov04}, which re-expresses the dynamics of a  BCS superconductor in terms of precessing spins. Using the language of Ref.~\onlinecite{Barankov04}, the non-equilibrium two time contour Green's functions 
\begin{align}
G^{\text{R}}_{\epsilon_k}(t,t')&=-i \Theta(t-t') \left\langle  \big [\Psi_k[t], \Psi_k^{\dagger}[t']\big ]_+\right \rangle,\\
G^{\text{K}}_{\epsilon_k}(t,t')&=-i  \left\langle  \big [\Psi_k[t], \Psi_k^{\dagger}[t']\big]_-\right \rangle,\\
G^{\text{A}}_{\epsilon_k}(t,t')&=i \Theta(t'-t) \left\langle \big  [\Psi_k[t], \Psi_k^{\dagger}[t']\big]_+\right \rangle,
\end{align}
can be rephrased in terms of a vector $\vec \Omega$, where the Keldysh Green's function in Nambu space at equal times takes the particularly simple form 
$
G^{\rm K}_\epsilon(t,t)=1+f_0^\epsilon{\bf {R}}_\epsilon^\dagger(t)\tau_3{\bf R}_\epsilon(t),
$     
and
$
\vec \Omega \cdot \vec \tau={\bf {R}}_\epsilon^\dagger(t)\tau_3{\bf R}_\epsilon(t),
$
where ${\bf \tau} =(\tau^1,\tau^2,\tau^3)^T$ and $\tau^i$ the Pauli matrices.
The dynamics of $\vec \Omega$ are determined by
\begin{equation}
\dot{ \vec\Omega}=2 \vec b_{\rm eff}(t) \times \vec \Omega,
\end{equation}
with $\vec b_{\rm eff}=(\Delta(t),0,\epsilon)^T$. Thus if $\vec \Omega$ follows the external ramp of $\Delta(t)$ (determining the effective field $\vec b_{\rm eff}(t)$) adiabatically so do the Green's functions, which fully characterize the non-equilibrium dynamics. It is easy to  check, that adiabatic evolution is guaranteed if 
\begin{equation}
\frac{1}{\epsilon T_0}\ll 1.\label{eq:BCS_ad}
\end{equation}
For a thermodynamic system $\epsilon$ can be arbitrarily small, and thus given any $T_0$ one can always find a part of the energy spectrum, which does not follow the deformation adiabatically. However, if $T_0$ is sufficiently large, the extent of this part of the energy spectrum becomes negligible. In this sense the evolution in the BCS-superconductor can be considered to be adiabatic if $T_0$ is such that the regime where $\frac{1}{\epsilon T_0}\ll 1$ is not fulfilled can be neglected. 

The condition on  adiabaticity Eq.~\eqref{eq:BCS_ad} is checked  in the top two rows of Fig.~\ref{fig:Blochs}, which show the dynamics of $\vec \Omega(t)$ (red lines) on the Bloch sphere as well as the adiabatic prediction (straight blue line) for different $\epsilon$ and $T_0$ (upper row $T_0=5$, lower row $T_0=10$).
As long as Eq.~\eqref{eq:BCS_ad} is fulfilled the adiabatic prediction agrees with the time evolved $\vec \Omega(t)$ at large times. In the regime where   Eq.~\eqref{eq:BCS_ad} is violated the vector $\vec \Omega(t)$ processes around the adiabatic prediction with constant radius $r=\Delta x=\Delta y$, where $\Delta x$ and $\Delta y$ are the deviations from the adiabatic prediction in the two directions orthogonal to it (see far upper right panel in Fig.~\ref{fig:Blochs}).
For the distribution of the quasi-modes (in Nambu space) $f(\epsilon)$ which fulfills $f(\epsilon)=\vec \Omega_3$ we can thus use the adiabatically deformed ensemble for all energies, which fulfill Eq.~\eqref{eq:BCS_ad}. This is shown in the bottom row of Fig.~\ref{fig:Blochs}. The three dimensional plot shows the distribution function in the perfectly adiabatically deformed ensemble ($T_0\to \infty$) as a blue line in dependency of $\epsilon$ compared to the result of the adiabatic deformation at finite $T_0$ ($T_0=1$ left and $T_0=5$ right) as red lines. The deviations from the adiabatic description $\Delta x$ and $\Delta y$ (as defined in the rightmost plot of the first row of Fig.~\ref{fig:Blochs}) are shown as circles around their mean value.

Next, we consider the limit $T_0\to \infty$, which means that the adiabatically deformed ensemble description is valid for any energy scale (at finite $T_0$ valid at energy scales $\epsilon\gg 1/T_0$). We then calculate the frequency $\omega$ dependent dirty limit optical conductivity $\sigma(\omega)$  by replacing the energy argument in the Fermi functions of Eqs. (3.9) and (3.10) in Ref.~\onlinecite{Mattis58} by the corresponding  adiabatic deformed one ($\epsilon\to \pm\sqrt{\epsilon^2-\Delta_0^2}$). In Fig.~\ref{fig:noneq_cond}(a) we show the real part of the optical conductivity calculated this way for initial temperature $T=0.3$ (blue dashed line) compared to the equilibrium optical conductivity at different $T$ (black to yellow lines). None of the thermal curves can be used to reproduce the adiabatically deformed one. The non-thermal additional in-gap content arises due to the fact that more energy is placed in the low energy modes compared to a thermal ensemble after deformation. The inset demonstrates that the superfluid stiffness $A$ is zero (blue line). The multi-colored line shows the equilibrium result. This means that the deformed BCS theory predicts an optical conductivity similar to the thermal one (with quantitatively changed line shape), but without the characteristic delta distribution (superfluid stiffness) contribution at $\omega=0$, which supports the super-current response characteristic to a superconductor.  In Fig.~\ref{fig:noneq_cond}(b) we show the optical conductivity for the adiabatically deformed ensemble for additional initial temperatures. At very low temperature the  strong divergence found in the conductivity for the adiabatically deformed ensemble mimics the behavior of the delta-distribution in the thermal case. Adiabatic passage thus leads to a quantitatively non-thermal behavior of the optical conductivity in a BCS superconductor strictly lacking superfluid stiffness.

\begin{figure}[t]
\centering
\includegraphics[width=\columnwidth]{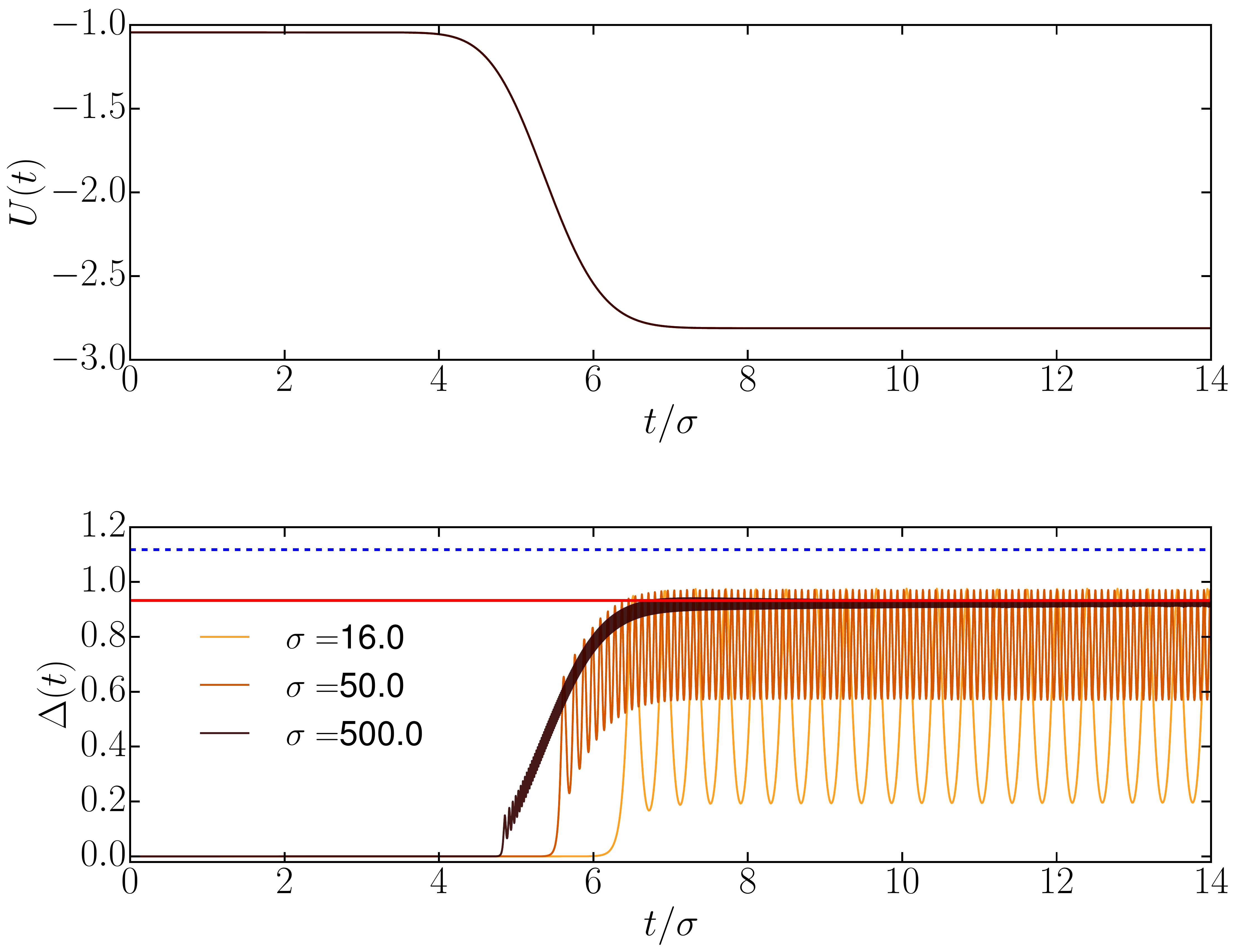}
\caption{Time evolution of the gap $\Delta(t)$ (bottom panel) for solving the self-consistent equation Eq.~\eqref{eq:Delta_SC} given the time dependent interaction $U(t)$ (top panel). $\sigma$ controls the adiabaticity of the $U(t)$ ramp. As $\sigma\to \infty$ the asymptotic behavior of $\Delta(t)$ is correctly described by the adiabatically deformed ensemble (red solid line), while a thermal description fails (blue dashed line).   }
\label{fig:ut_ad}
\end{figure}

As a final example we also consider the case where a time dependent interaction $U(t)$ is given and the gap $\Delta(t)$ is determined by the self-consistent equation\cite{Barankov04,Kennes17} 
\begin{equation}
\Delta(t)=U(t)\sum_{k}f_{0}(k)Tr\left[\tau^{-}\mathbf{R}_\epsilon^{\dagger}(t)\tau_{3}\mathbf{R}_\epsilon(t)\right].
\label{eq:Delta_SC}
\end{equation}   
We concentrate on a three dimensional model of a cubic lattice with nearest neighbor hopping. Thus we study a featureless semi-elliptic density of states and set units by choosing the bandwidth to be $4$.
We choose a temperature of $T=0.1$ and the profile of $U(t)$ following
\begin{equation}
U_0  + \frac{A}{4} \left[  {\rm{erf}}\left( \frac{t -\delta}{\sigma}\right)  + {\rm{erf}} \left(\frac{\delta}{\sigma}\right) \right]^2
\end{equation}
Therefore $\sigma$ controls the speed $v\sim 1/\sigma$ of the interaction ramp, where $\sigma\to\infty$ is the adiabatic limit.
Representative results are summarized in Fig.~\ref{fig:ut_ad}. The upper panel shows the interaction ramp for $U_0=-1.045$, $A=-1.77$ and $\delta=5\sigma$. The lower panel shows the self-consistently determined gap for different values of $\sigma$. The oscillatory behavior reminiscent of the abrupt quench,\cite{Barankov04} decrease in height as $\sigma$ is increased. The oscillations are a consequence of  the residual precession of the spin analyzed in Fig.~\ref{fig:Blochs} at large times, if the process is not sufficiently adiabatic.
For increasing $\sigma$ the oscillations vanish and the steady value is well described by the prediction of the adiabatically deformed ensemble (red line), while a thermal description of the system does not yield convincing results (blue dashed line). The adiabatically deformed ensemble  thus provides  a significantly simpler route to the steady state than performing the transient time evolution. 



\section{Majorana Chain}
As a final example we study the topological Kitaev chain \cite{Kitaev01}
\begin{equation}
H=\sum\limits_{j=1}^L\left(-c(t)c_j^\dagger c_{j+1}+Mc_j c_{j+1}-\frac{m(t)}{2}c_j^\dagger c_{j}+{\rm H.c.}\right)\;\;.
\end{equation} 
We choose periodic boundary conditions and for convenience $|M|=1$. Using Nambu vectors $\Psi_k=(c_k,c^\dagger_{-k})^T$ the Hamiltonian can be written as $H=\sum_k \Psi_k^\dagger H_k \Psi_k$ with
$
H_k=\frac{\Delta_k}{2}{\bf n}_k\cdot {\bf \sigma} 
$
and
\begin{align}
{\bf n}_k=\frac{2}{\Delta_k}(0,-\sin\, k, -m(t)+c(t)\cos\, k )\\
\Delta_k=2\sqrt{(-m+c\;\cos\,k)^2+\sin^2\, k},
\end{align}
where ${\bf \sigma} =(\sigma^x,\sigma^y,\sigma^z)^T$ and $\sigma^i$ the Pauli matrices. 
We concentrate on adiabatic deformations in the regime $c>m$, which at $T=0$  displays topological order and the Berry phase $\arg\left[\cos(\pi\omega_1)\right]$ is $\pi$, with
\begin{equation}
\omega_1=\frac{1}{2\pi}\oint\left(\frac{\partial_k n_k^i}{n_k^j}\right)dk \;\;\;\;\;\;\;\;i\neq j.
\end{equation}

\begin{figure}[t]
\centering
\includegraphics[width=0.9\columnwidth]{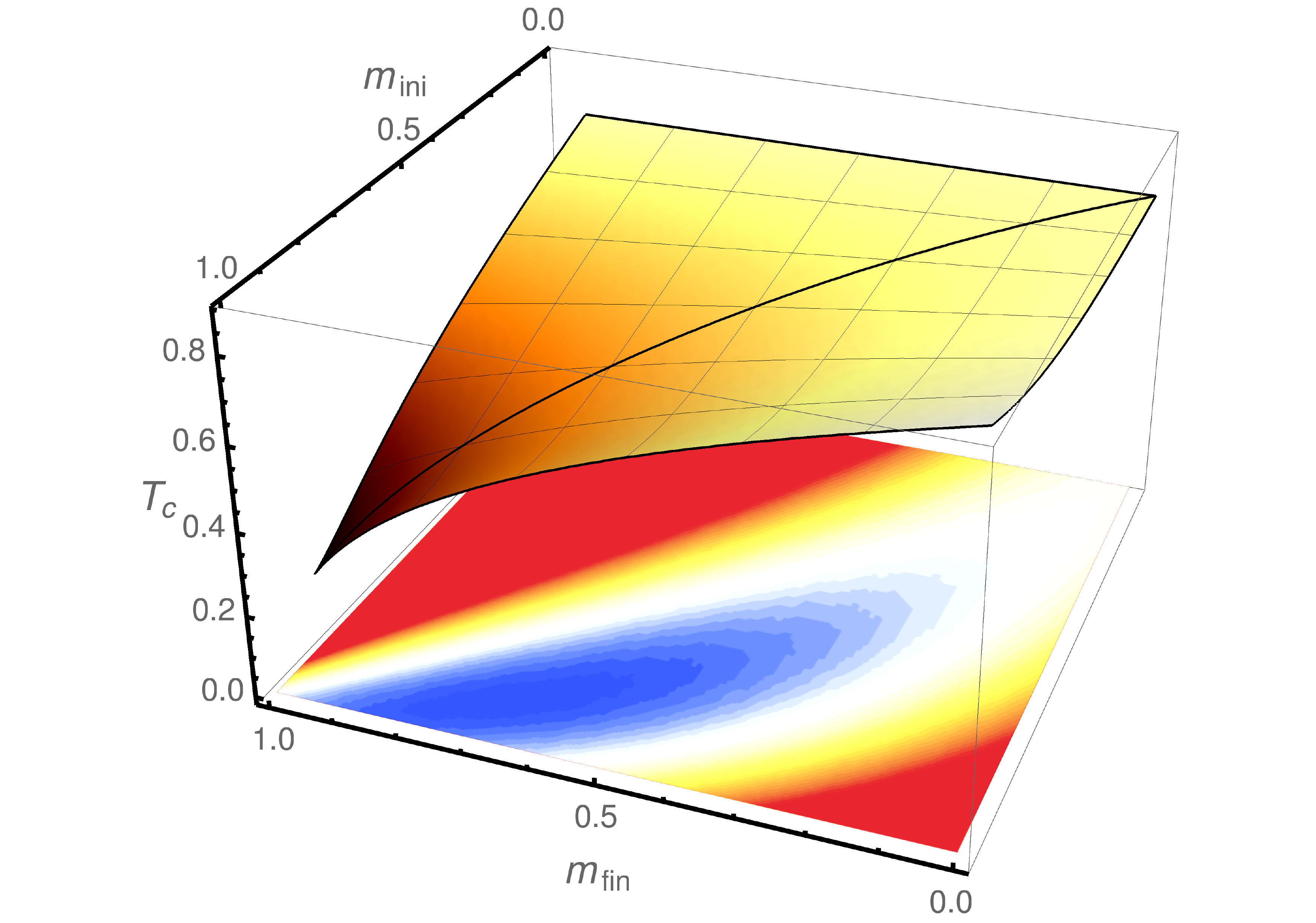}
\caption{ Critical temperature obtained from adiabatically varying $m(t)$ from $m_{\rm ini}$ to $m_{\rm fin}$ at fixed $c(t)=1$. The thick diagonal line indicates the cut $c_{\rm ini}=c_{\rm fin}$, where we recover the equilibrium result. The lower contour plot shows $T_c-T_c^{\rm eq}$, where $T_c^{\rm eq}$ is the equilibrium critical temperature w.r.t. $c_{\rm fin}$ on a diverging color map (white, blue, red colors indicate a change of $0$, negative, positive magnitude, respectively). }
\label{fig:topo2}
\end{figure}

To probe the topological properties for mixed states we use the generalization of the Berry phase,\cite{Uhlmann86,Viyuela14} 
\begin{equation}
\Phi_{\rm U}=\arg\left\{\cos(\pi\omega_1)\cos\left[\oint\left(\frac{\partial_k n_k^i}{2n_k^j}\right){\rm sech}\left(\frac{\Delta_k}{2T}\right)dk\right]\right\}.
\label{eq:Phi_therm}
\end{equation} 
Like the Berry phase at $T=0$, $\Phi_{\rm U}=0$ ($\Phi_{\rm U}=\pi$) in the trivial (topological) phase with a sharp transition separating these two phase at $T=T_c$.
As we exclusively concentrate on cases that show a finite gap $\Delta_k$, the evolution  must be slow on the scale of the gap $v \ll \min(\Delta_k)$ to ensure adiabatic deformation. Adiabatic deformation from a set $(c_{\rm ini},m_{\rm ini})\to (c_{\rm fin},m_{\rm fin})$ modifies $\Phi_{\rm U}$ to  
\begin{align}
&\Phi_{\rm U}=\arg\bigg\{\cos(\pi\omega_1(c_{\rm fin},m_{\rm fin}))\notag\\
&\times\cos\left[\oint\left(\frac{\partial_k n_k^i(c_{\rm fin},m_{\rm fin})}{2n_k^j(c_{\rm fin},m_{\rm fin})}\right){\rm sech}\left(\frac{\Delta_k(c_{\rm ini},m_{\rm ini})}{2T}\right)dk\right]\bigg\}.\label{eq:Phi_ad}
\end{align} 
Therefore, for the adiabatically deformed ensemble the previously thermal distribution condensed in the argument $\frac{\Delta_k}{2T}$ of the ${\rm sech}$ in Eq.~\eqref{eq:Phi_therm} has to replaced by a ${\rm sech}$ with a different (non-thermal) argument $\frac{\Delta_k(c_{\rm ini},m_{\rm ini})}{2T}$. This non-thermal distribution in turn modifies $T_c$. Results for the critical temperature $T_c$ when adiabatically varying $m(t)$ are summarized in Fig.~\ref{fig:topo2}. Adiabatic deformation thus opens a route to alter the critical temperature, not due to cooling, but due to the non-thermal distribution function (condensed in the argument of the ${\rm sech}$) in Eq.~\eqref{eq:Phi_ad}.

%

\section{Summary}
We propose adiabatic deformation of (thermal) ensembles to engineer non-thermal states of matter exhibiting largely unexplored physics.
The speed with which this adiabatic passage has to be undertaken depends on the system under scrutiny. A case by case study should be conducted to study, whether an adiabatically deformed ensemble can be achieved before residual couplings to the environment spoil the closed-ness of the quantum system and drive it back into a thermal state. 
We performed four case studies for simple but important systems: (a) a single oscillator, (b) a dimerized finite one-dimensional chain of spinless fermions,\cite{Gadway15,Meier16} (c) a BCS superconductor with time dependent gap-function as well as (d) a Majorana chain. All show that the adiabatically deformed ensemble harbors interesting physics inaccessible by thermal pathways. 
For the dimerized chain we show that long ranged correlations can be achieved.  This could open a route to study Luttinger liquid behavior in systems for which accessing low temperatures is difficult.  For the BCS-type superconductor we investigated the optical conductivity, one of the experimentally most relevant observables. We find a clearly non-thermal behavior  most prominently reflected in a line shape qualitatively similar to the thermal one but lacking a superfluid stiffness. For a topological system we find that due to the deformation the critical temperature at which the topological phase is lost can be increased by the non-thermal nature of the distribution function after deformation. 
Future research should address whether the designed non-thermal states presented here can harbor unknown (hidden) phases inaccessible by thermal equilibrium.

{\it{Acknowledgements}}--- I thank Christoph Karrasch, Volker Meden and Andrew Millis for many fruitful discussions and helpful comments. D.M.K. was supported by Deutsche Forschungsgemeinschaft Grant No. KE 2115/1-1. Simulations were performed with computing
resources granted by RWTH Aachen University under project
rwth0013. \\

{}

\end{document}